\documentclass[aps,prb,twocolumn,superscriptaddress,longbibliography]{revtex4-2}
\usepackage{graphicx}
\usepackage{bm}% bold math
\usepackage{natbib}
\usepackage{color}
\usepackage{epstopdf}
\usepackage{amsmath}
\usepackage{url}
\usepackage{amssymb}
\definecolor{darkGreen}{RGB}{0,110,0}
\definecolor{darkBlue}{RGB}{5,4,170}
\usepackage[colorlinks=true, urlcolor=darkBlue,citecolor=darkBlue,linkcolor=darkBlue]{hyperref}
\usepackage{pifont}
\usepackage{gensymb}

\begin{document}

\title{Edge states, Majorana fermions and topological order in superconducting wires with generalized boundary conditions}

\author{Alfonso Maiellaro}
\affiliation{Dipartimento di Ingegneria Industriale, Universit\`{a} degli Studi di Salerno, Via Giovanni Paolo II, 132, I-84084 Fisciano (SA), Italy}
\author{Francesco Romeo}
\affiliation{Dipartimento di Fisica "E.R. Caianiello", Universit\`{a} degli Studi di Salerno, Via Giovanni Paolo II, 132, I-84084 Fisciano (SA), Italy}
\author{Fabrizio Illuminati}
\affiliation{Dipartimento di Ingegneria Industriale, Universit\`{a} degli Studi di Salerno, Via Giovanni Paolo II, 132, I-84084 Fisciano (SA), Italy}
\affiliation{INFN, Sezione di Napoli, Gruppo collegato di Salerno,Italy}

\date{July 2, 2022}

\begin{abstract}
We study the properties of one-dimensional topological superconductors under the influence of generic boundary conditions mimicking the coupling with external environments. We identify a general four-parameters classification of the boundary effects and show that particle-hole and reflection symmetries can be broken or preserved by appropriately fixing the boundary parameters. When the particle-hole symmetry is broken, the topological protection of the edge modes is lost due to the hybridization with the external degrees of freedom (quasiparticle poisoning). We assess the robustness of the edge modes in the various regimes by considering different quantifiers of topological properties. In particular, we investigate the resilience of the long-distance, edge-to-edge quantum mutual information and squashed entanglement, measuring the nonlocal correlations of the Majorana excitations. Besides their relevance for the open dynamics of topological systems, these results may provide a useful guide to the appropriate embedding of low-dimensional topological systems on nanodevices in realistic conditions.
\end{abstract}

\maketitle

\section{Introduction}
\label{introduction}
In the last two decades, topological states of matter and topological materials have been subject to intense investigation, both for their importance in the understanding of fundamental quantum matter and their potential role in technological applications \cite{PhysRevLett.61.2015,PhysRevLett.90.206601,Yu61,PhysRevB.61.10267,RevModPhys.82.3045,Sato_2017,PhysRevB.84.195442}. In particular, topological superconductors, hosting non-Abelian edge states known as Majorana zero modes (MZMs), are considered among the most promising systems for the implementation of fault-tolerant topological quantum computers \cite{pachos_2012,PhysRevLett.111.203001,doi:10.1146/annurev-conmatphys-030212-184337}, as different experiments have provided preliminary evidence of MZMs and their signatures \cite{Mourik1003,Nadj-Perge602}.

The theoretical idea of realizing and store quantum information using MZMs dates back to the seminal work by Alexei Kitaev  \cite{Kitaev_2001}. In his pioneering contribution, Kitaev proposed a model describing a one-dimensional spinless $p$-wave superconductor able to sustain unpaired Majorana excitations. These modes, protected by the particle-hole symmetry \cite{Sato_2017}, localize at the edges of the system and maintain topological robustness against local perturbations. Several proposals have followed on such footsteps in order to identify different condensed matter platforms for the engineering of $p$-wave pairing. Key ingredients to realize an emerging $p$-wave pairing are the superconducting proximity effect between an $s$-wave superconductor and the surface of a topological insulator \cite{PhysRevLett.100.096407} or, more conveniently, the superconducting proximization of semiconducting heterostructures. In the latter case, platforms with strong spin-orbit coupling such as InAs and InSb nanowires \cite{PhysRevLett.104.040502}, semiconducting thin films with broken time-reversal symmetry \cite{PhysRevB.81.125318} and one-dimensional semiconducting nanowires with Rashba spin-orbit coupling and Zeeman field effect \cite{PhysRevLett.105.077001,PhysRevLett.105.177002} have been proposed.

Boundary conditions play a relevant role in determining the presence or the absence of MZMs. Indeed, topological systems with periodic boundary conditions (PBCs) cannot feature unpaired Majorana modes, while for systems with open boundary conditions (OBCs), topologically protected modes nucleate at the system edges in topologically ordered phases. When applicable, the bulk-edge correspondence \cite{PhysRevResearch.2.013147} assures that knowledge of a system with PBCs provides information on the presence of topological modes at the edges of the corresponding system with OBCs. Besides particle-hole symmetry, a relevant role in such correspondence is played by the reflection symmetry that takes a coordinate $x$ into $-x$. The topological correspondence between systems with OBCs and PBCs holds when homogeneous couplings are involved, thus ensuring that the system with OBCs is invariant under coordinate reflection. Situations also exist in which the bulk-edge correspondence cannot be invoked. Indeed, robustness of the MZMs has been demonstrated both in the presence of disorder effects \cite{PhysRevB.84.144526,PhysRevLett.107.196804,PhysRevB.88.064506,Neven_2013,PhysRevB.89.144506,PhysRevB.94.115166,PhysRevB.98.035142,PhysRevB.92.195119,PhysRevLett.110.146404,PhysRevLett.106.057001,PhysRevB.83.155429,PhysRevB.71.245124,PhysRevLett.105.046803,PhysRevB.86.205135} and in models with long-range hopping terms \cite{MaiellaroGeoFrust,PhysRevB.97.041109,PhysRevB.88.165111,PhysRevLett.113.156402,PhysRevB.95.195160,MaielTopTie}, two situations in which the bulk-edge correspondence does not apply.

The presence of Majorana fermions is typically revealed by signatures that are characteristic of topologically ordered phases \cite{PhysRevLett.101.120403,PhysRevLett.103.107002,PhysRevB.81.085101,PhysRevLett.109.227001,PhysRevB.87.024515,PhysRevB.96.054504,PhysRevB.96.201109}. In particular, the presence of Majorana bound states can be detected, for instance, by tunnelling spectroscopy \cite{PhysRevLett.98.237002,PhysRevLett.103.237001,PhysRevB.63.144531,nano9060894,PhysRevB.86.224511,PhysRevLett.119.136803,PhysRevLett.110.126406,PhysRevLett.109.267002} or by using interferometric devices able to identify the anomalous $4 \pi$-periodic Josephson effect \cite{Kwon,PhysRevB.84.180502}.

The above detection methods are particularly relevant in the context of complex geometries involving normal-superconductor (NS) interfaces \cite{PhysRevB.87.165414,PhysRevB.84.180509,PhysRevB.91.024514,PhysRevLett.105.077001,PhysRevB.86.180503,PhysRevB.86.220506,PhysRevB.91.214513,PhysRevB.82.041405}. In these structures two terminals, playing the role of source and drain, are coupled to the system of interest in order to directly monitor the transition to a topologically ordered phase \cite{doi:10.1021/nl303758w,Nishio_2011,doi:10.1021/nl203380w}. Under this condition, the effect of coupling of the topological system with the source/drain electrodes needs to be modeled and investigated by considering boundary conditions different from the hard-wall confinement associated to OBCs. In fact, hard-wall confinement is neither desirable nor possible in realistic experimental conditions, and thus generic boundary conditions have to be considered. The Kitaev-type modelling of such contexts incorporates information on the external environment via generalized boundary conditions. Imposing generalized boundary conditions is expected to provide a consistent picture of open topological systems whenever non-equilibrium current flows are either absent or can be considered negligible.

Thus motivated, in the present work we investigate the physics of Kitaev chains in the presence of generic boundary conditions. We derive a general four-parameters classification of the boundary effects and demonstrate that particle-hole and reflection symmetries can either be broken or be preserved depending on the values taken by the boundary parameters. We investigate systematically the robustness of the edge modes under the influence of the different boundary conditions by using several estimators, including the long-distance edge-to-edge entanglement measuring the nonlocal correlation of the Majorana excitations \cite{MaieIllum1,MaieIllum2}.

The paper is organized as follows. In Sec. \ref{analyticalCalculations} we introduce the tight-binding equations of a Kitaev chain coupled to proximized electrodes under the influence of generalized boundary conditions, and we show that particle-hole and reflection symmetries can be broken by appropriately setting the boundary terms without directly acting on the chain parameters. In Sec. \ref{NumericalResults} we report relevant results of various numerical studies in which, by looking at different physical quantities, we investigate systematically the conditions for the presence of MZMs. Among the quantities investigated, we show that a special role is played by the topological entanglement between the system edges. In Sec. \ref{conclusions} we draw our conclusions and discuss some future outlook, while in Appendices \ref{AppA} and \ref{AppB} we report on the the technical details and the mathematical tools used throughout the paper.

\section{Kitaev chain with generalized boundary conditions}
\label{analyticalCalculations}
The Kitaev chain model describes spinless fermions subject to a p-wave superconducting pairing and constrained to move along a one-dimensional lattice. The time evolution of the on site fermionic wave function $\psi_{n}$ is described by the equation:
\begin{eqnarray}
\label{BdGKC}	
i \hbar \partial_t \psi_n=-\mu \sigma_z \psi_n &-& t \sigma_z (\psi_{n+1}+\psi_ {n-1})+\nonumber\\
&+& i \Delta \sigma_y (\psi_{n+1}-\psi_{n-1}),
\end{eqnarray}
where $\sigma_{x,y,z}$ denote the Pauli matrices and $n \in \{1, \dots, L\}$ specifies the position along the chain. The tight-binding parameters $t$, $\Delta$, $\mu$ define, respectively, the nearest-neighbour hopping, the superconducting pairing and the chemical potential. The Nambu spinor $\psi_n=(u_n\ \ v_n)^T$ provides information about the quasiparticle weight so that $u_n$ and $v_n$ are the particle and hole components of the wave function at the $n-th$ lattice site. The OBCs hard-wall confinement constraint amounts to impose $\psi_{0}=\psi_{L+1}=0$ at the two chain ends in Eq. (\ref{BdGKC}). The above choice is appropriate in order to investigate the topological properties of an isolated Kitaev chain under the associated bulk-edge correspondence. On the other hand, hard-wall confinement only seldom models correctly the actual experimental conditions; in most cases, either unavoidable or desired couplings with the environment have to be taken into account; as a result, one needs to impose generic boundary conditions. In fact, the Kitaev chain model supplemented with deformed boundary conditions precisely incorporates information on the external environment. This method, which shares similarities with a self-energy approach proposed in Ref. \cite{Aguadoself}, allows to investigate an open topological system when the non-equilibrium effects induced by a current flow are either negligible or absent. The latter condition is surely met when the electrodes-system hybridization dominates over other concurrent effects.
\begin{figure*}
	\includegraphics[scale=0.45]{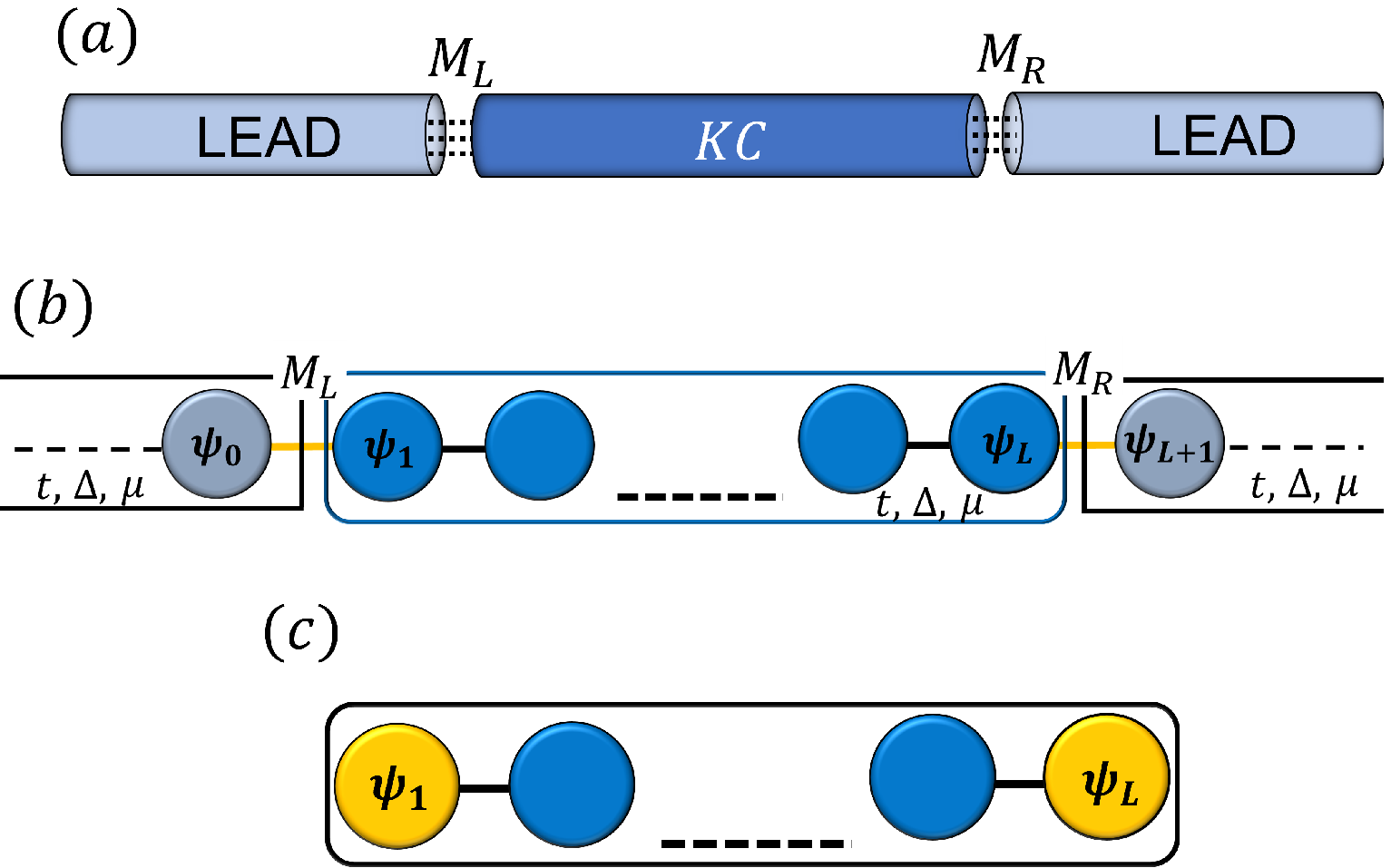}
	\hspace{0.5cm}
	\includegraphics[scale=0.52]{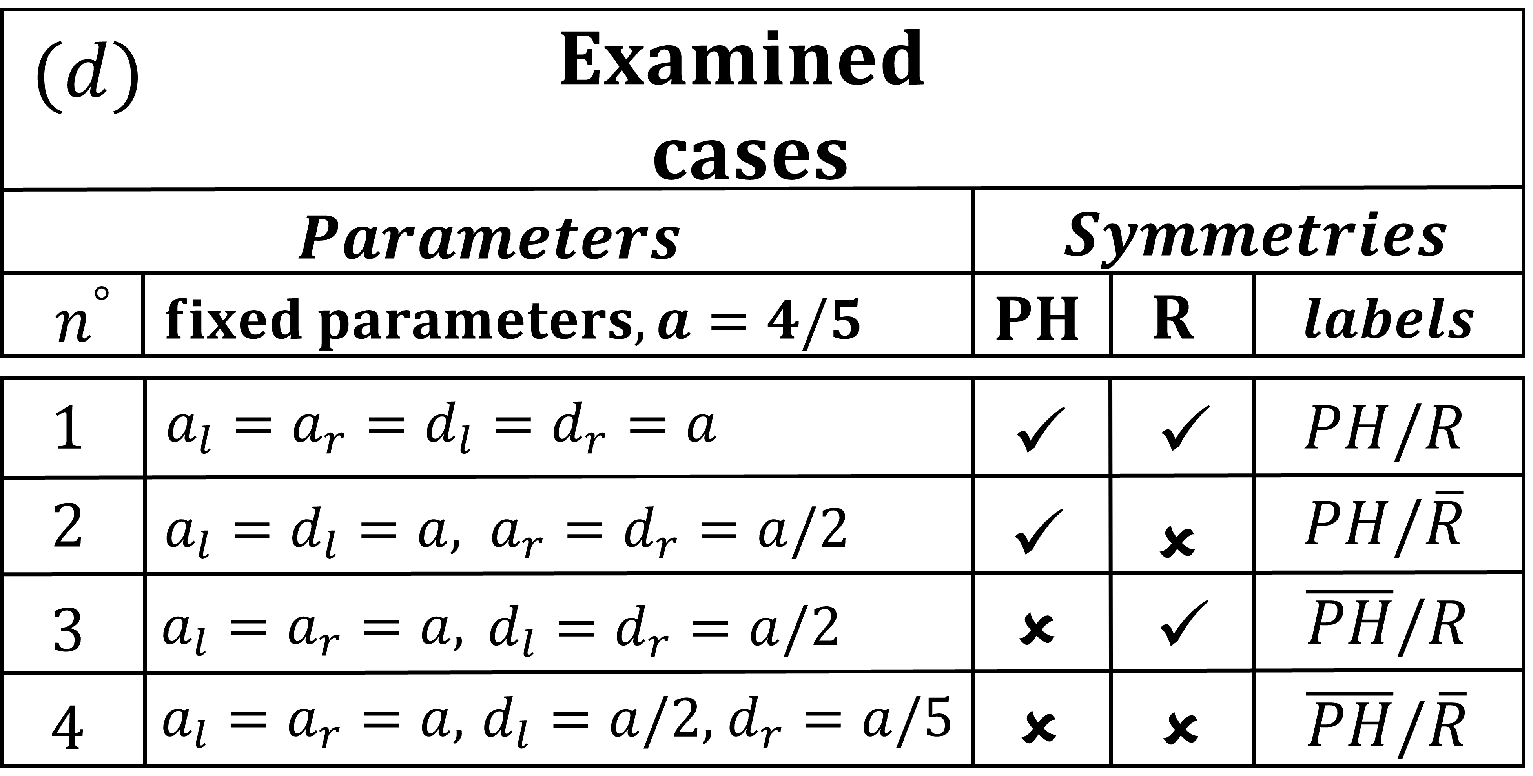}
	\caption{Panel (a): Scheme of a Kitaev chain coupled to an environment modeled by a pair of superconducting leads. Panel (b): Tight-binding representation of the system. Panel (c): Effective chain model with generalized boundary conditions. Table (d): A paradigmatic set of generalized boundary conditions. Breaking or conservation of the particle-hole ($PH$) and reflection ($R$) symmetry is determined by the choice of the pertinent boundary parameters. The $a$ parameter is fixed at the reference value $a=4/5$.}
	\label{Scheme}
\end{figure*}
The coupling of the Kitaev chain with the environment can be addressed by means of generalized boundary conditions, namely $\psi_0=M_l \psi_1$ and $\psi_{L+1}=M_r \psi_L$, which depend on environmental wave functions $\psi_0$ and $\psi_{L+1}$. The coupling between the sites belonging to the environment, labelled by $n=0$ and $n=L+1$, and the two ends of the chain is illustrated in panel (b) of Fig. \ref{Scheme}. The boundary properties are completely described by the auxiliary matrices $M_l$ and $M_r$. In general, the latter are unknown \textit{a priori}, even though their structure can be at least partially determined by imposing suitable physical constraints. In order to implement such constraints, let us introduce generic complex-valued matrices in particle-hole space:
\begin{eqnarray}
     M_l=\left(
     \begin{array}{cc}
     	a_l&b_l\\
     	c_l&d_l\\
     \end{array}
     \right),&&\ \ M_r=\left(
     \begin{array}{cc}
     	a_r&b_r\\
        c_r&d_r\\
     \end{array}
     \right).
     \label{MatricesBC}
\end{eqnarray}
By resorting to such generalized boundary conditions in Eq. (\ref{BdGKC}), we obtain a set of deformed tight-binding equations. They differ from those of the Kitaev chain with OBCs by a local renormalization of the edge potentials. These boundary effects are clearly evident by considering the deformed equations for the wave functions $\psi_1$ and $\psi_L$ of two chain ends:
\begin{eqnarray}
\label{tbB}
i\hbar \partial_t \psi_{1}&=&-(\mu \sigma_{z}+N_{l})\psi_{1}-t\sigma_{z}\psi_{2}+i \Delta \sigma_{y}\psi_{2} \, , \\
i\hbar \partial_t \psi_{L}&=&-(\mu \sigma_{z}+N_{r})\psi_{L}-t\sigma_{z}\psi_{L-1}-i \Delta \sigma_{y}\psi_{L-1} \, . \nonumber
\end{eqnarray}
We see that the dynamics depends on the boundary potentials $-N_l=-(t \sigma_z+i \Delta \sigma_y)\ M_l$ and $-N_r=-(t \sigma_z-i \Delta \sigma_y)\ M_r$ which vanish when OBCs are considered (i.e. when $M_{l/r}=0$). The fermionic statistics dictates that $N_l$ and $N_r$ be diagonal matrices in particle-hole space and thus Eq. (\ref{MatricesBC}) takes the form:
\begin{eqnarray}
	M_l=\left(
	\begin{array}{cc}
		a_l&-\frac{\Delta}{t}d_l\\
		-\frac{\Delta}{t}a_l&d_l\\
	\end{array}
	\right),&&\ \ M_r=\left(
	\begin{array}{cc}
	a_r&\frac{\Delta}{t}d_r\\
	\frac{\Delta}{t}a_r&d_r\\
	\end{array}
	\right).
	\label{Matrices2}
\end{eqnarray}\\
A further constraint comes from conservation of probability, which is expressed by the continuity equation, $\partial_t \rho_n+J_{n+1}-J_{n}=0$, relating the probability density $\rho_n=|u_n|^2+|v_n|^2$ and the probability current density
\begin{eqnarray}
	J_n=\frac{2t}{\hbar}\ Im\biggl[\psi^\dagger_{n-1} (\sigma_z-i\ \frac{\Delta}{t}\ \sigma_y)\psi_n\biggr].
	\label{currentDensity}
\end{eqnarray}
In the stationary regime, $J_n$ is a conserved quantity. In particular, when non-equilibrium effects induced by a current flow are negligible or absent, $J_n=0$ for each site index $n$. This condition is fulfilled both by the model with generalized boundary conditions and by the Kitaev chain with OBCs; this allows an appropriate comparison between systems with OBCs and systems with generalized boundary conditions. Exploiting Eq. (\ref{currentDensity}) with $n=1$ and $n=L+1$, we get:
\begin{eqnarray}
	\begin{split}
&J_1 \propto \biggl(1-\frac{\Delta^2}{t^2}\biggr) \biggl[|u_1|^2 Im\bigl[a_l^*\bigr]-|v_1|^2 Im\bigl[d_l^*\bigr] \biggr]\\
&J_{L+1} \propto \biggl(1-\frac{\Delta^2}{t^2}\biggr) \biggl[|u_L|^2 Im\bigl[ a_r \bigr]-|v_L|^2 Im\bigl[ d_r \bigr] \biggr].
\end{split}
\label{Current2}
\end{eqnarray}
Since $J_1$ and $J_{L+1}$ have to vanish for any choice of parameters and for any quantum state of the chain, it follows that $M_l$ and $M_r$ are real-valued matrices with general expression given in Eq. (\ref{Matrices2}). Accordingly, the boundary potentials, see Eq. (\ref{tbB}), are related to the real-valued matrices as follows:
 \begin{eqnarray}
	N_l=\epsilon \left(
	\begin{array}{cc}
		a_l&0\\
		0&-d_l\\
	\end{array}
	\right),&&\ \ N_r=\epsilon \left(
	\begin{array}{cc}
		a_r&0\\
		0&-d_r\\
	\end{array}
	\right),
	\label{MatricesN}
\end{eqnarray}
where $\epsilon=(t^2-\Delta^2)/t$. Crucially, by acting on the form of $N_l$ and $N_r$, it is possible to selectively preserve reflection and particle-hole symmetry. Reflection symmetry is lost when $N_l \neq N_r$, while, depending on the choice of parameters, particle-hole symmetry can be preserved or broken. In particular, particle-hole symmetry is preserved for traceless boundary potentials (i.e. $Tr[N_l]=Tr[N_r]=0$).

\begin{figure*}
	\includegraphics[scale=0.25]{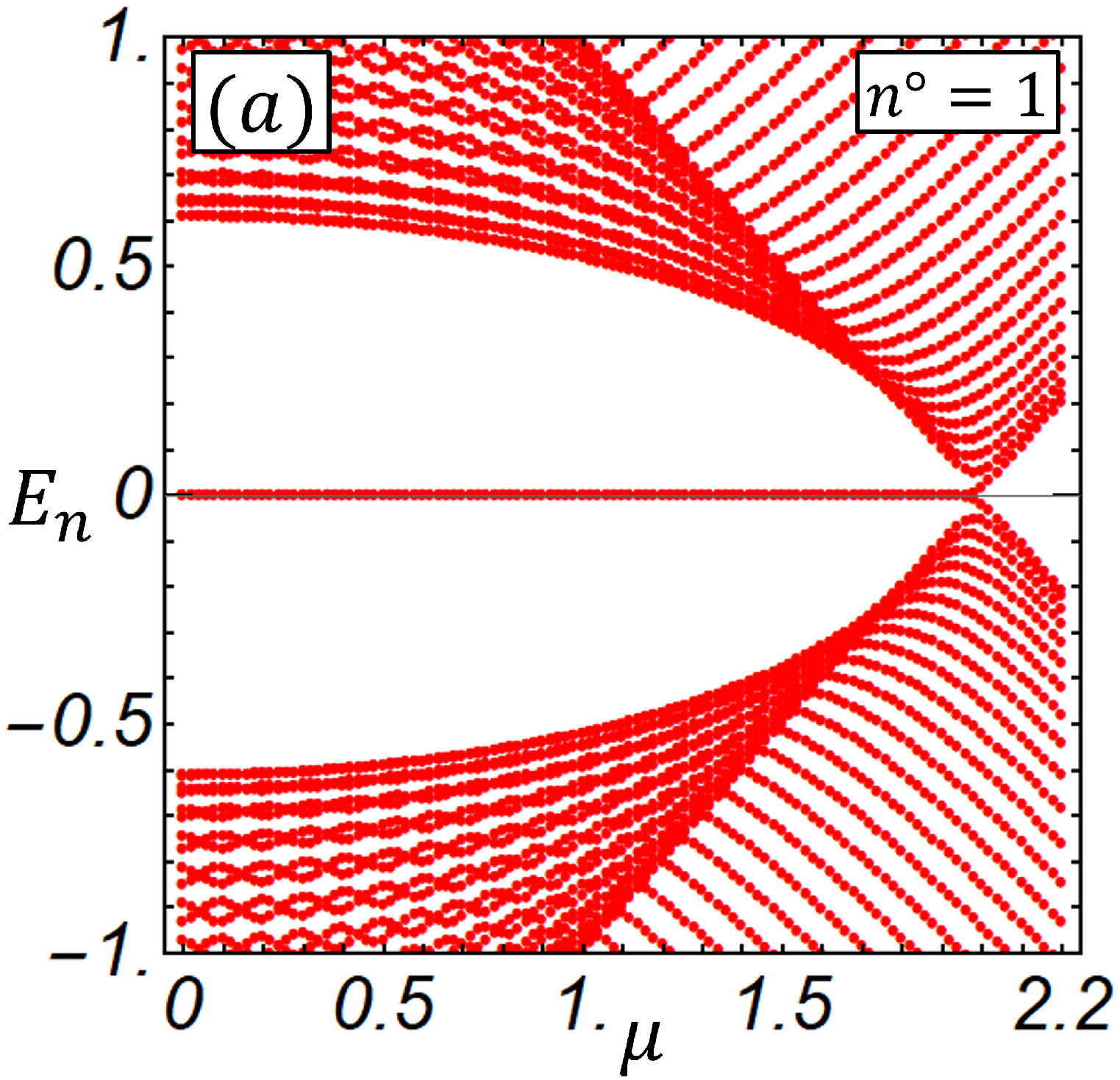}
	\hspace{0.05cm}
	\includegraphics[scale=0.25]{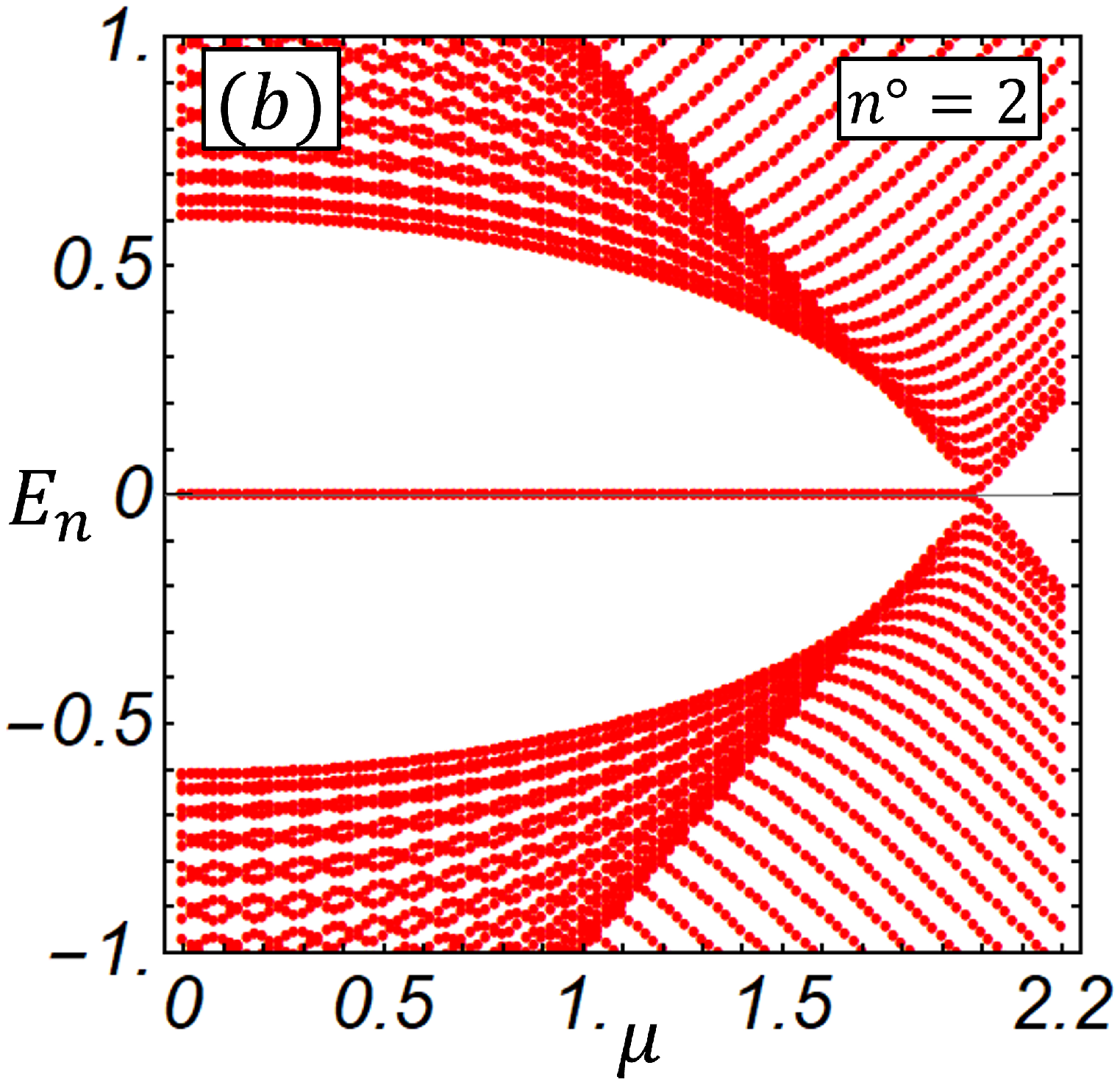}
	\hspace{0.05cm}
	\includegraphics[scale=0.25]{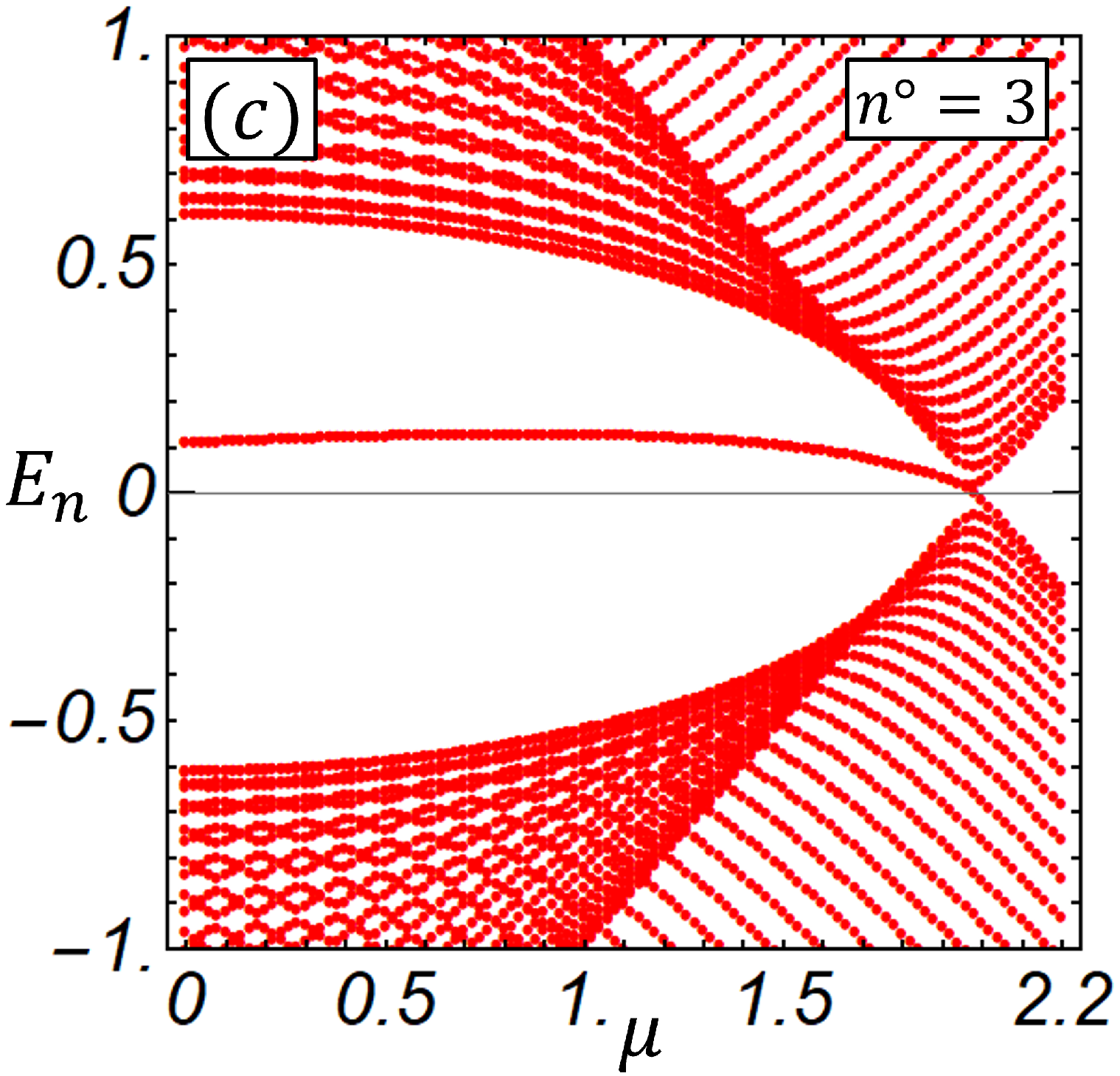}
	\hspace{0.05cm}
	\includegraphics[scale=0.25]{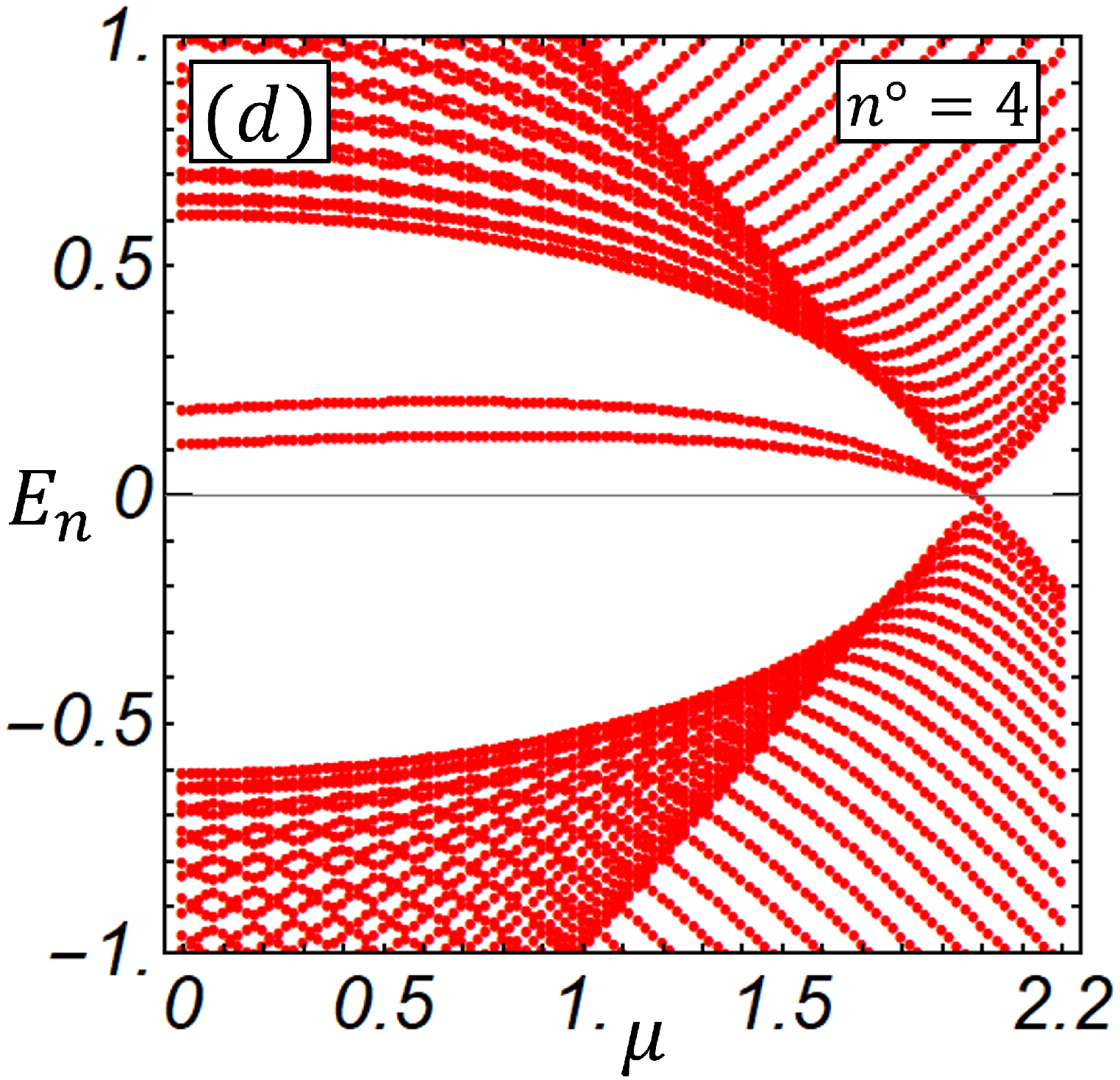}\\
	\vspace{0.3cm}
	\includegraphics[scale=0.25]{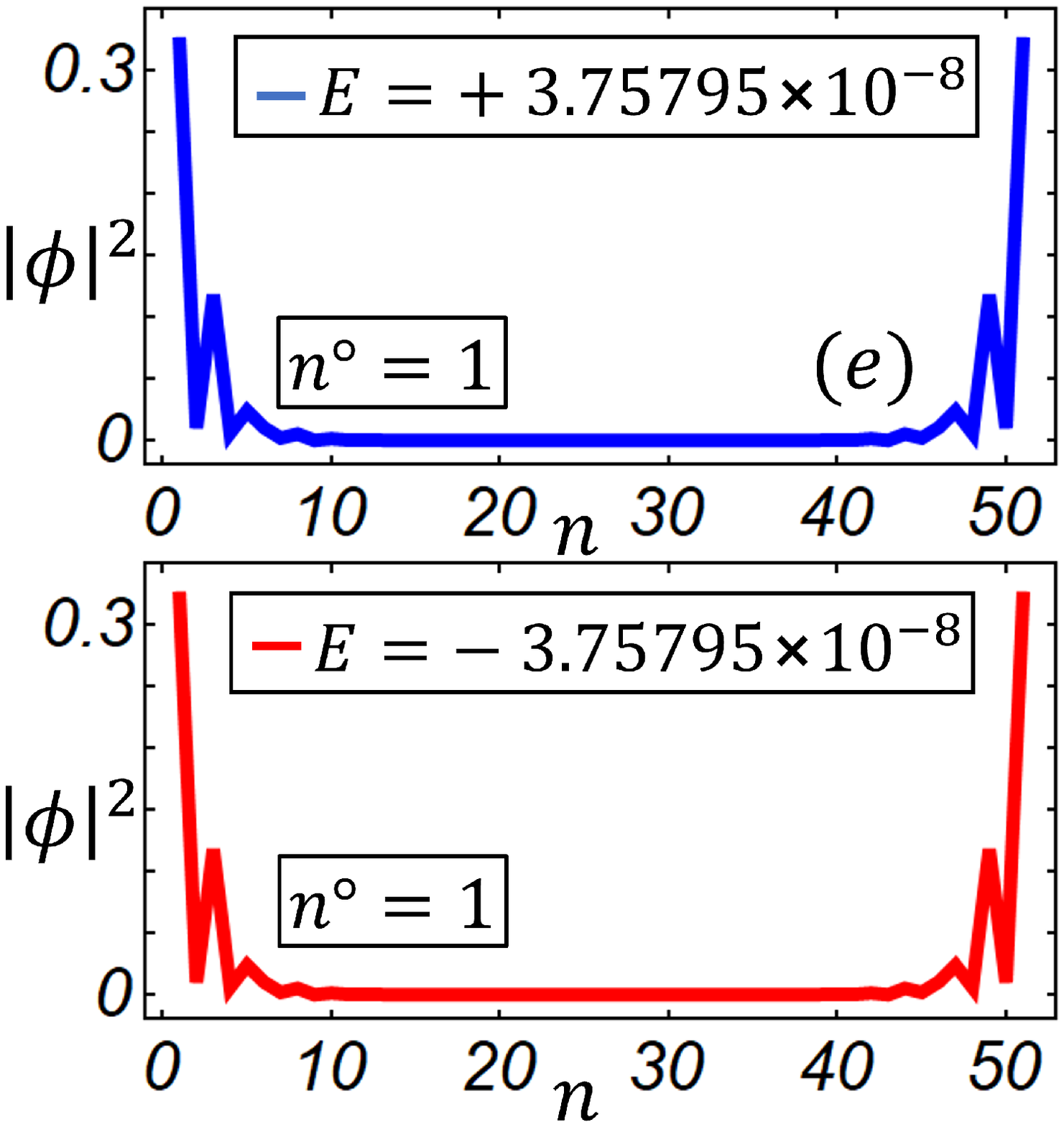}
	\hspace{0.05cm}
	\includegraphics[scale=0.25]{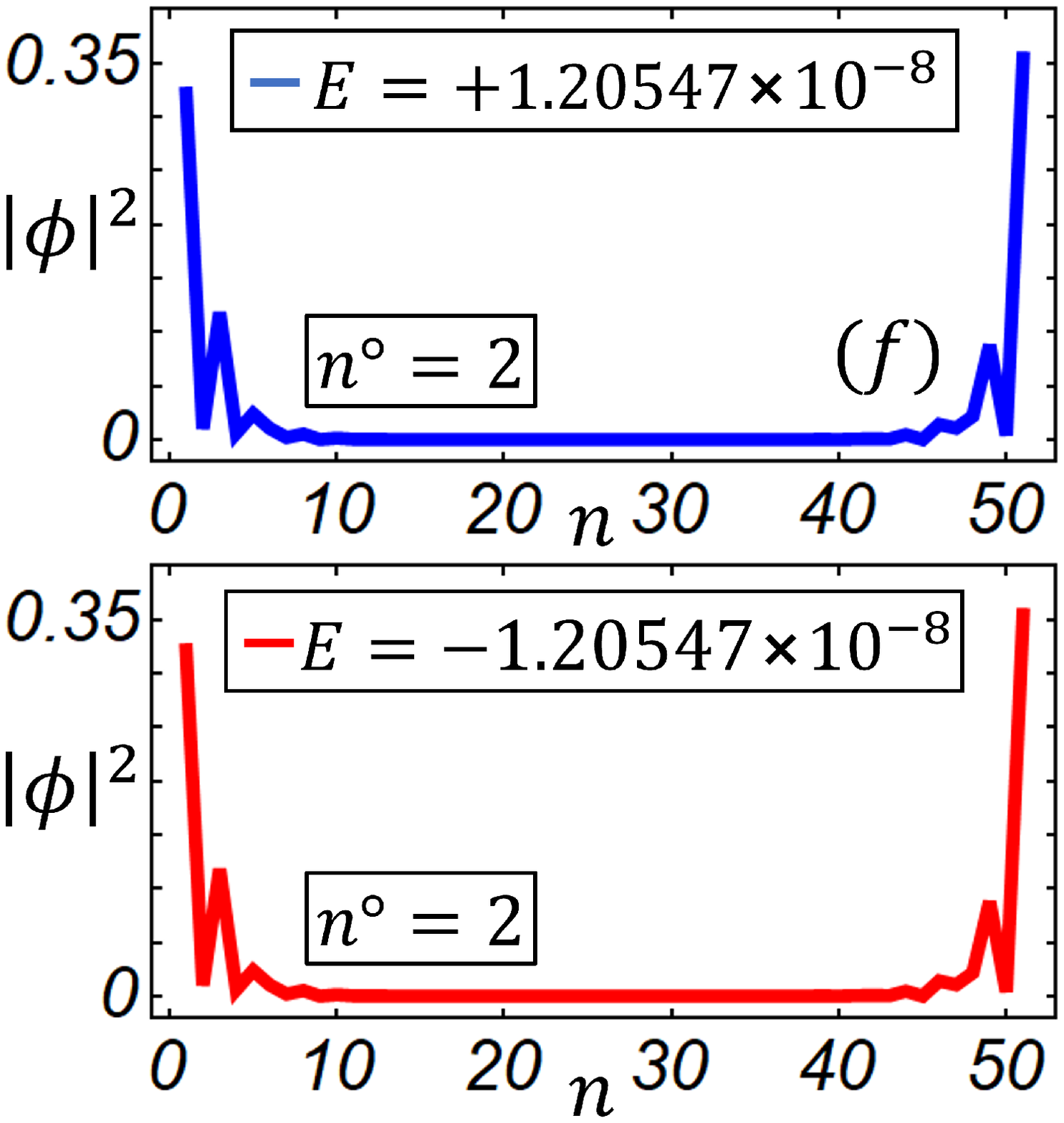}
	\hspace{0.05cm}
	\includegraphics[scale=0.25]{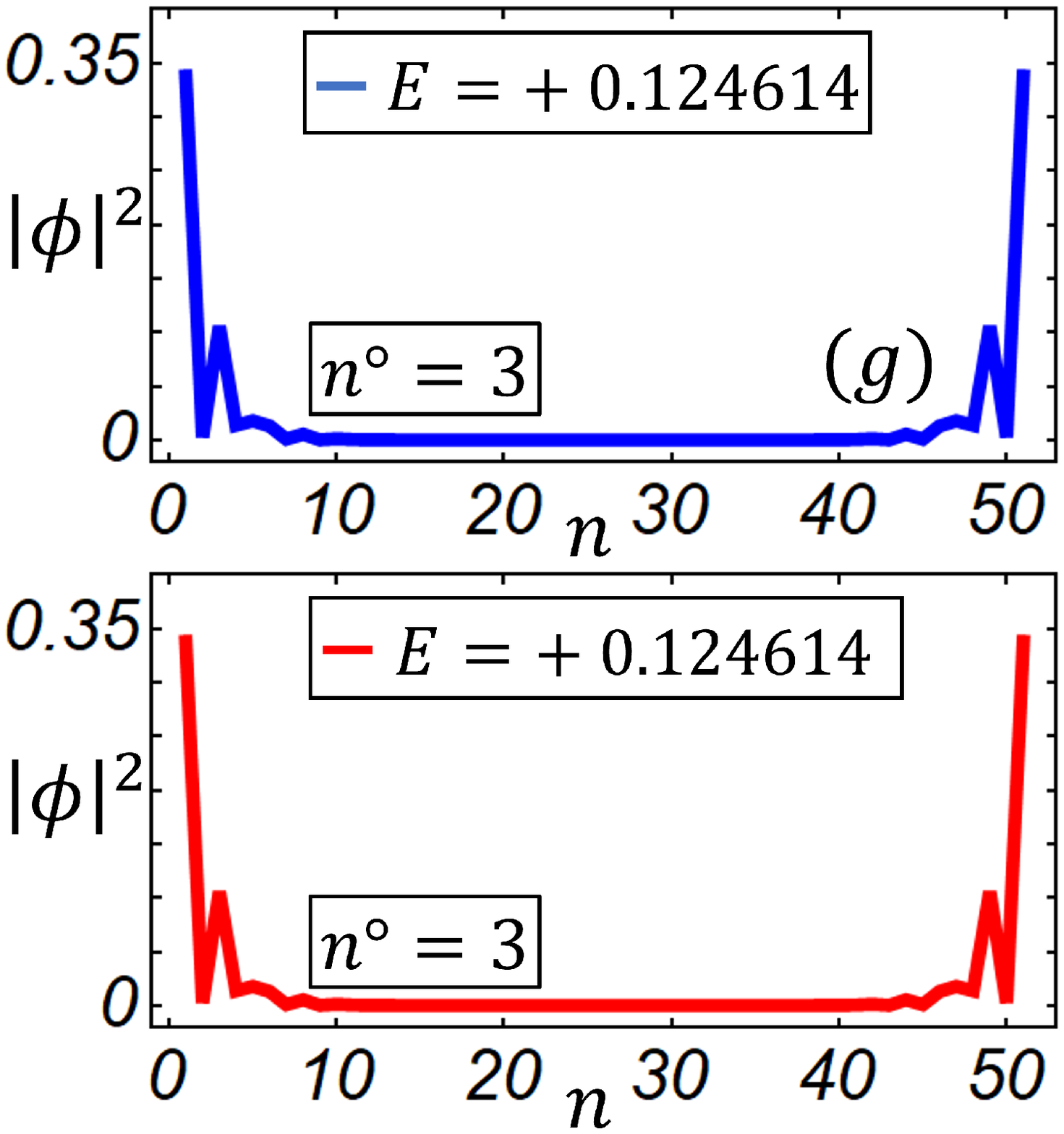}
	\hspace{0.05cm}
	\includegraphics[scale=0.25]{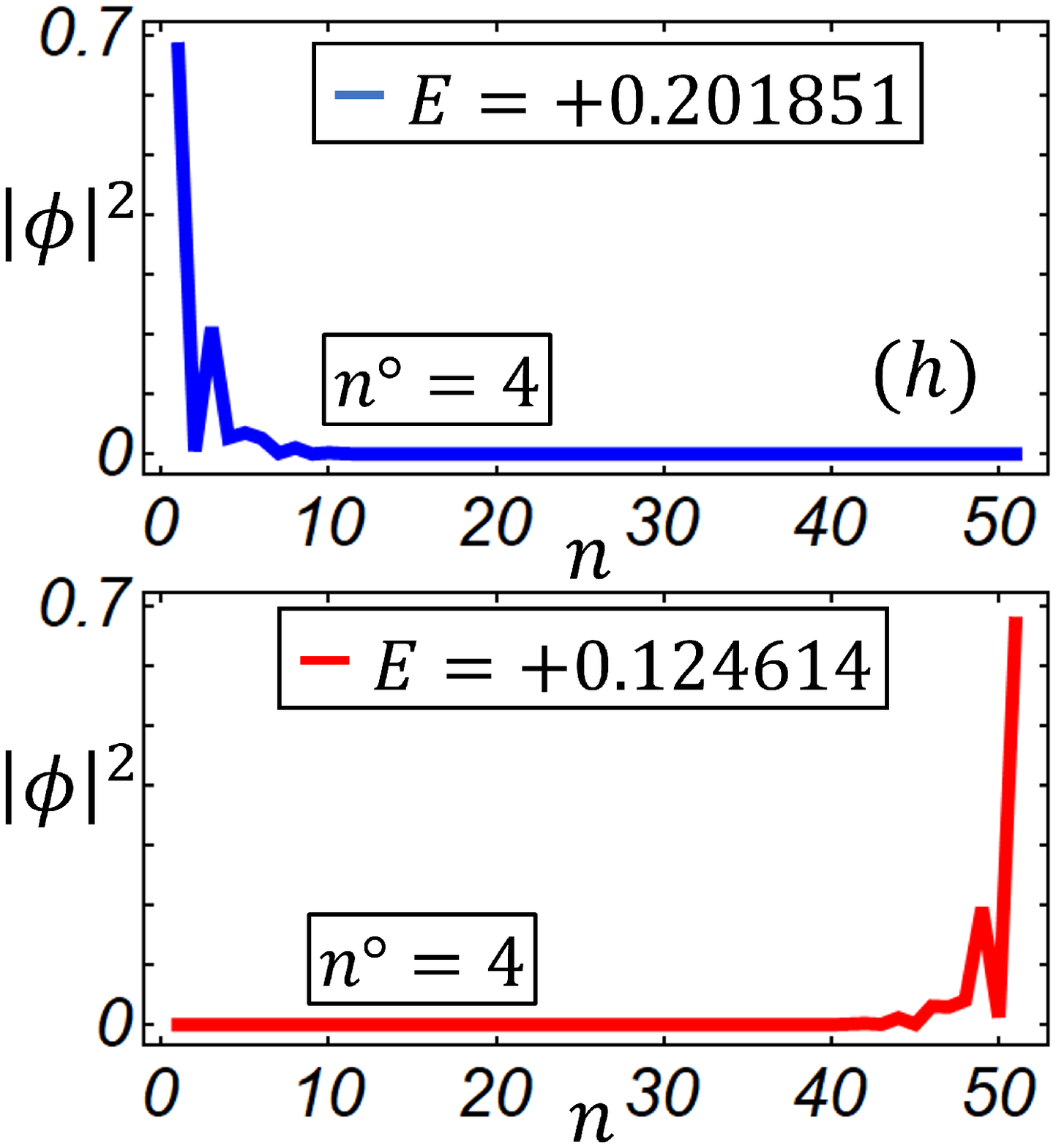}
	\caption{Panels (a) to (d): Energy spectrum of a Kitaev chain with generalized boundary conditions, as a function of the chemical potential $\mu$. Panels (e) to (h): Spatial spread of the sub-gap modes at fixed chemical potential $\mu=0.5$. The energy eigenvalues pertaining to each sub-gap mode are reported inside the inset. Each panel refers to the boundary conditions reported in table (d) of Fig. \ref{Scheme}. The remaining parameters have been fixed as: $t=1$, $\Delta=0.3$ and $L=51$.}
	\label{SpectralProperties}
\end{figure*}

Particle-hole symmetry is a fundamental consequence of any non-interacting mean field theory of superconductivity. For this reason, it would be tempting to impose that boundary potentials preserve this symmetry. However, experimental evidence supports the existence of mechanisms responsible for the breaking of this symmetry. Among them, we mention the quasiparticle poisoning, which is a non-interacting mechanism, and the ubiquitous electron-boson interactions in superconductors \cite{PHSb}. Thus, when the effective model of a Kitaev chain coupled to an environment is considered, generalized boundary conditions with particle-hole symmetry breaking cannot be neglected. In this way we obtain a four-parameters (symmetry-based) classification of the boundary effects.

\section{Results}
\label{NumericalResults}
Hereafter, we present a comprehensive analysis of the fate and robustness of topological order in a Kitaev chain model subject to generalized boundary conditions. In view of the high dimensionality of parameter space, we focus on four choices of the parameters, see the table in panel (d) of Fig. \ref{Scheme}, identified by the labels $n \degree 1,..., n \degree 4$ and belonging to distinct symmetry classes. Each choice is specified by setting appropriate values of $a_{l,r}$ and $d_{l,r}$ and corresponds to the breaking or conservation of the particle-hole and/or the reflection symmetry. In this way we can identify four symmetry classes, namely $PH/R$ ($n \degree 1$), $PH/\overline{R}$ ($n \degree 2$), $\overline{PH}/R$ ($n \degree 3$) and $\overline{PH}/\overline{R}$ ($n \degree 4$). Importantly, once a specific symmetry class has been assigned, we have verified that the results do not feature qualitative differences with those obtained with a different choice of parameters within the same class.

A prominent feature related to the topological order is the presence of zero-energy modes localized at the system boundaries. In order to monitor this signature,  in Fig. \ref{SpectralProperties}, panels (a)-(d), we show the evolution of the energy spectrum of the chain as a function of the chemical potential $\mu$. We observe that zero-energy modes appear for $\mu \leq 2t$ when the particle-hole symmetry is preserved (panels (a) and (b)). On the other hand, the energy of such modes becomes finite and positive when particle-hole symmetry is broken (panels (c) and (d)). In the latter case, an exact degeneracy of the sub-gap modes persists, see panel (c), until the reflection symmetry is also broken, see panel (d). Positive sub-gap energy levels, see panels (c) and (d), are induced by the positive sign of the $a$ parameter; we have also verified that the sign of the sub-gap energy levels is inverted if we consider negative values of the $a$ parameter and suitable values of the chemical potential, as reported in Appendix \ref{AppB}. At a closer inspection of panels (c) and (d), we see that particle-hole symmetry breaking mainly affects the sub-gap energy levels, while the bulk-like band structure remain substantially unaffected.

The modes profile is shown in panels (e) through (h) of Fig. \ref{SpectralProperties} (e)-(h) at fixed chemical potential $\mu = t/2$. For all symmetry classes, one invariably obtains localized modes with tails decaying inside the bulk. Moreover, the decay of the edge modes inside the bulk is quite insensitive to the symmetry class considered. A closer analysis shows that the modes profile loses definite parity when reflection symmetry is broken, as shown in panels (f) and (h). Furthermore, the simultaneous breaking of the particle-hole and reflection symmetry, see panel (h), completely disentangles the sub-gap modes which appear to be localized at different edges of the chain. Despite their localized character, the latter are non-topological modes strongly hybridized with the environment.

\begin{figure*}
	\includegraphics[scale=0.275]{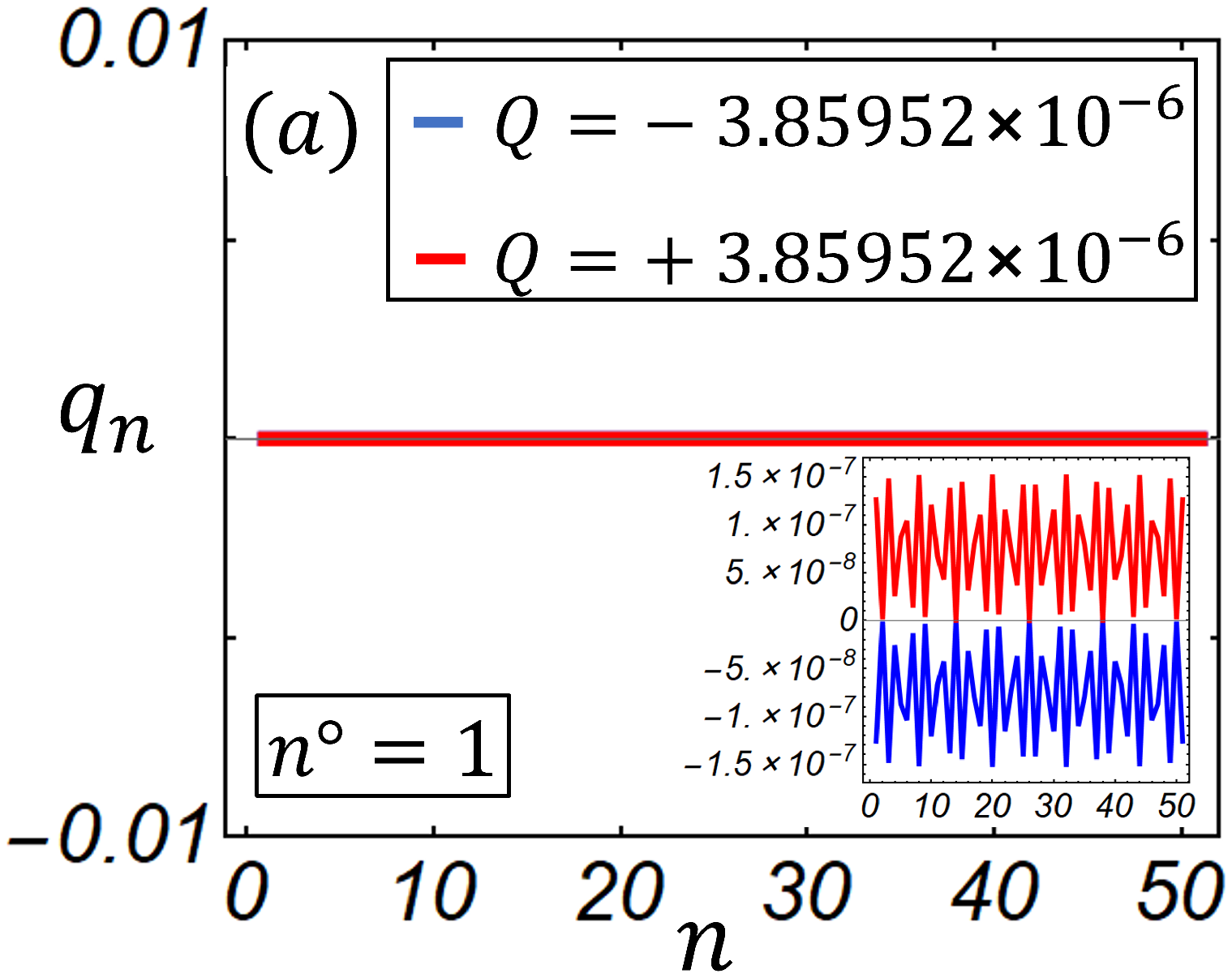}
		\hspace{0.02cm}
	\includegraphics[scale=0.275]{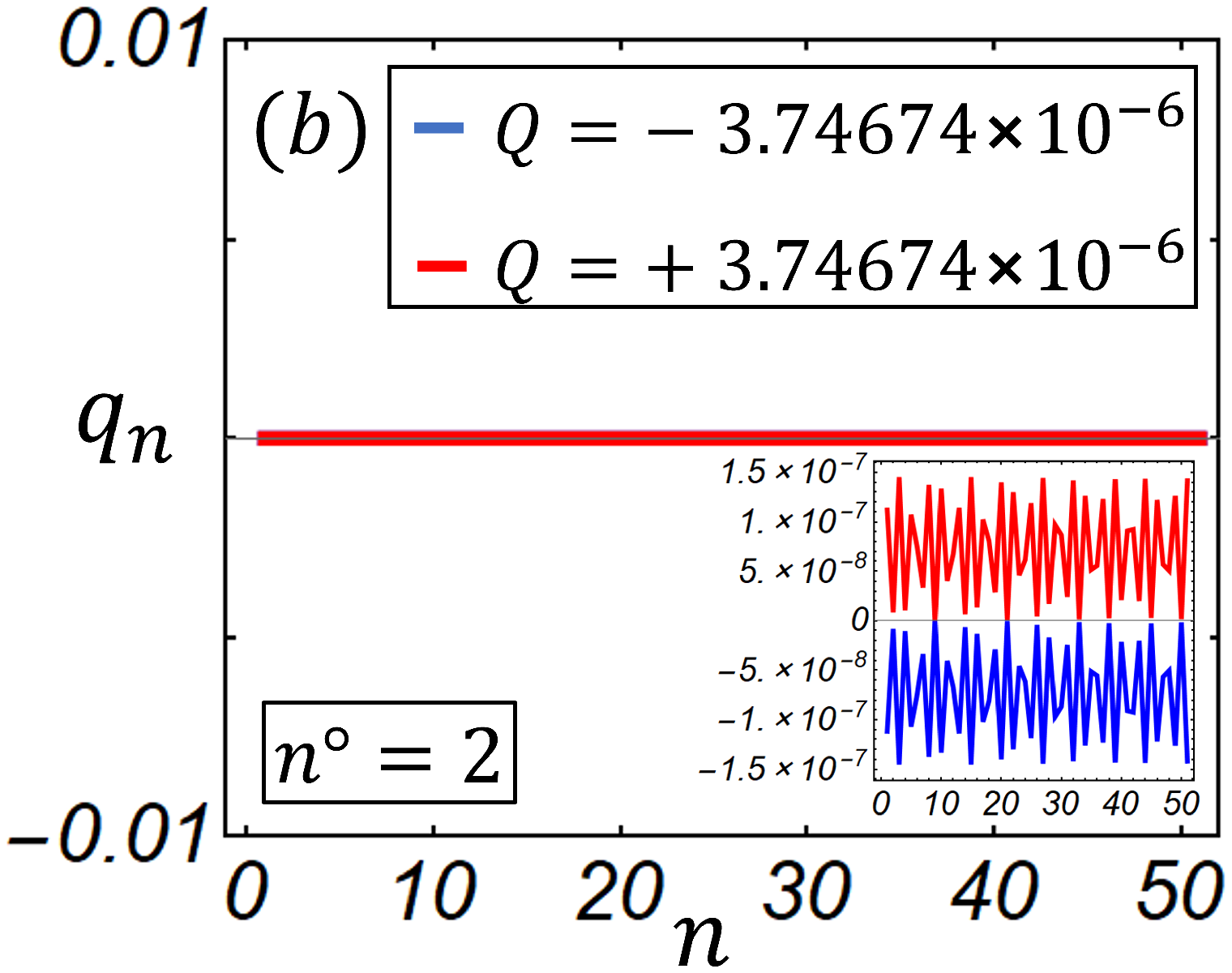}
	\includegraphics[scale=0.282]{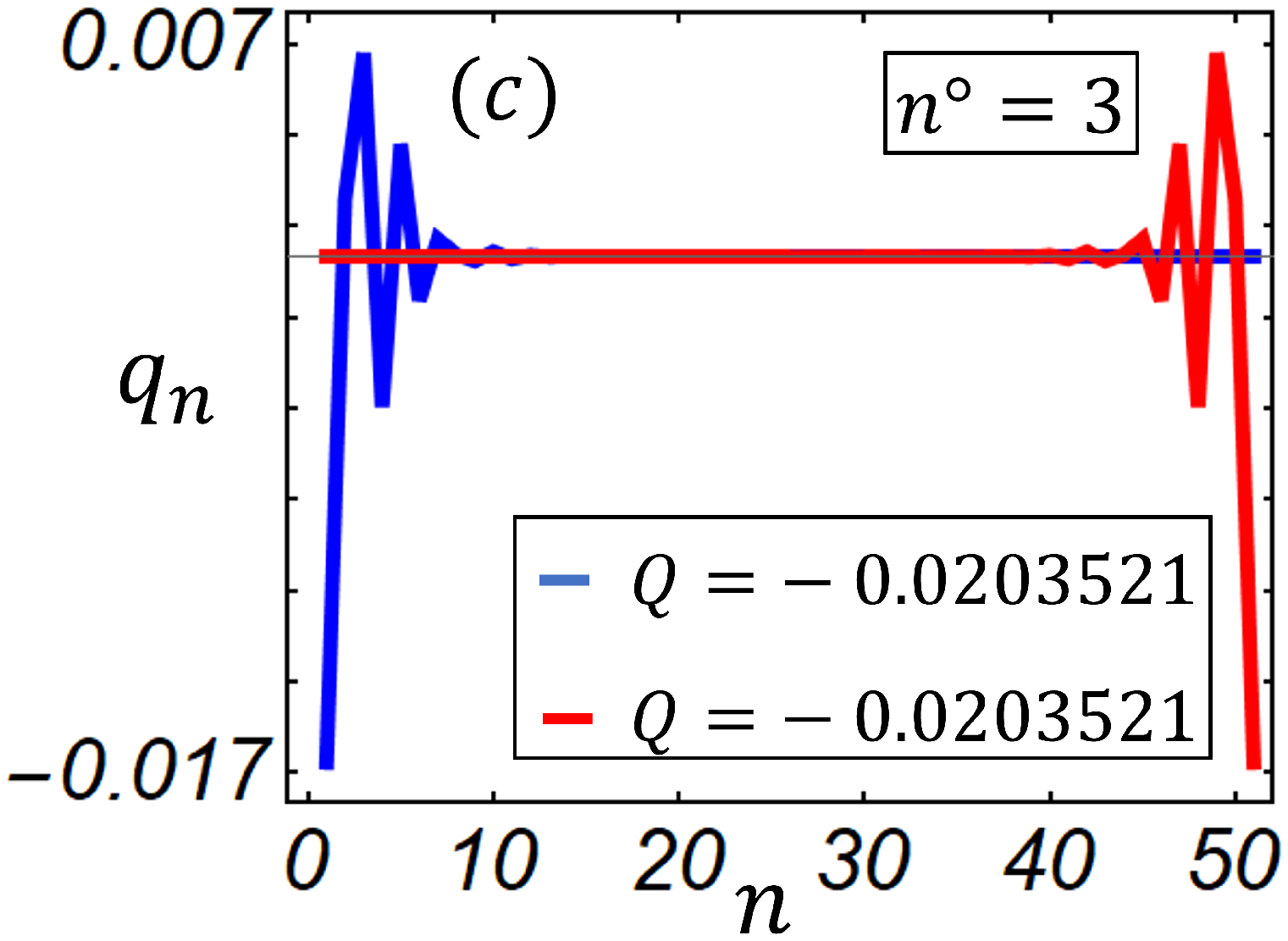}
	\includegraphics[scale=0.275]{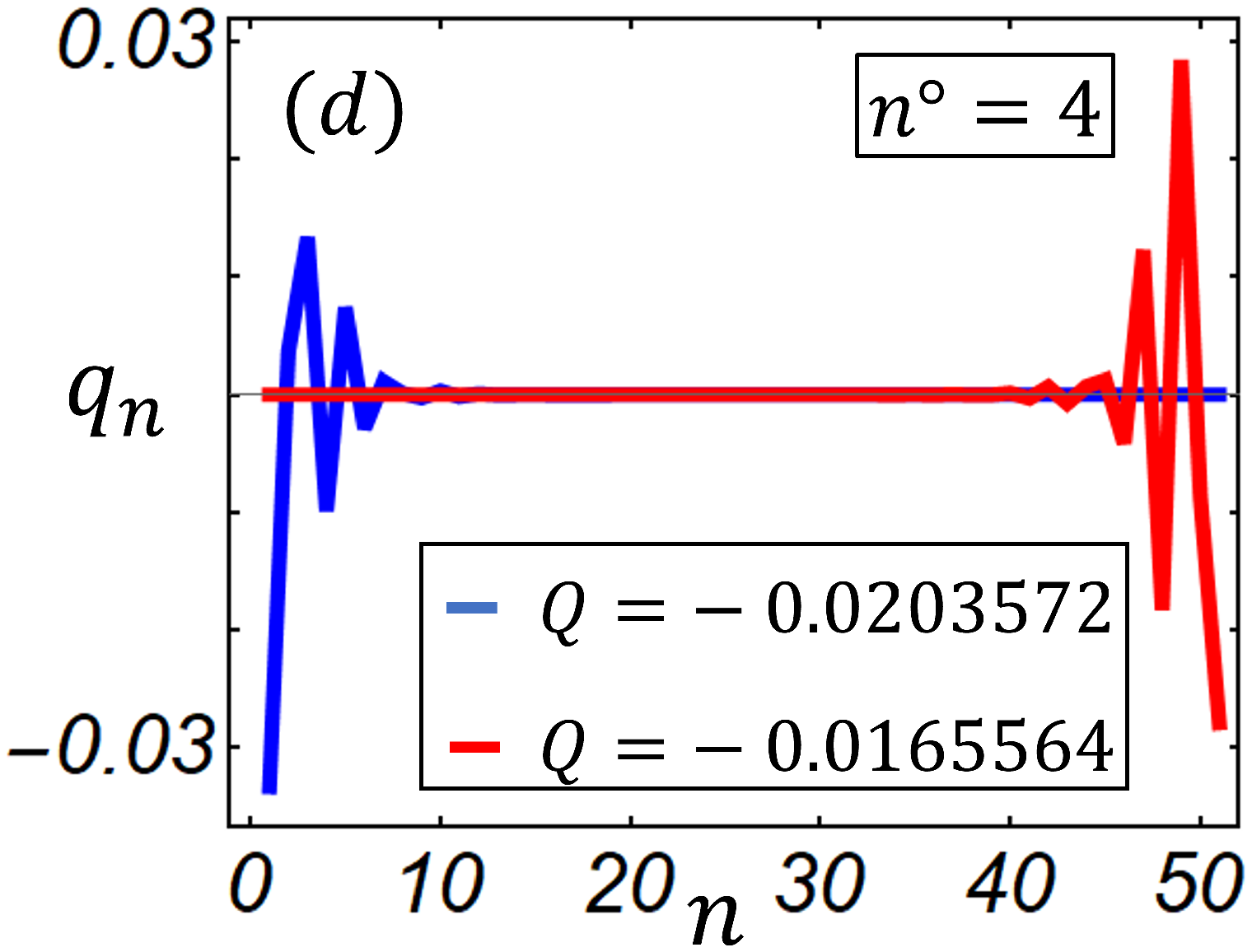}\\
		\vspace{0.3cm}
	\includegraphics[scale=0.27]{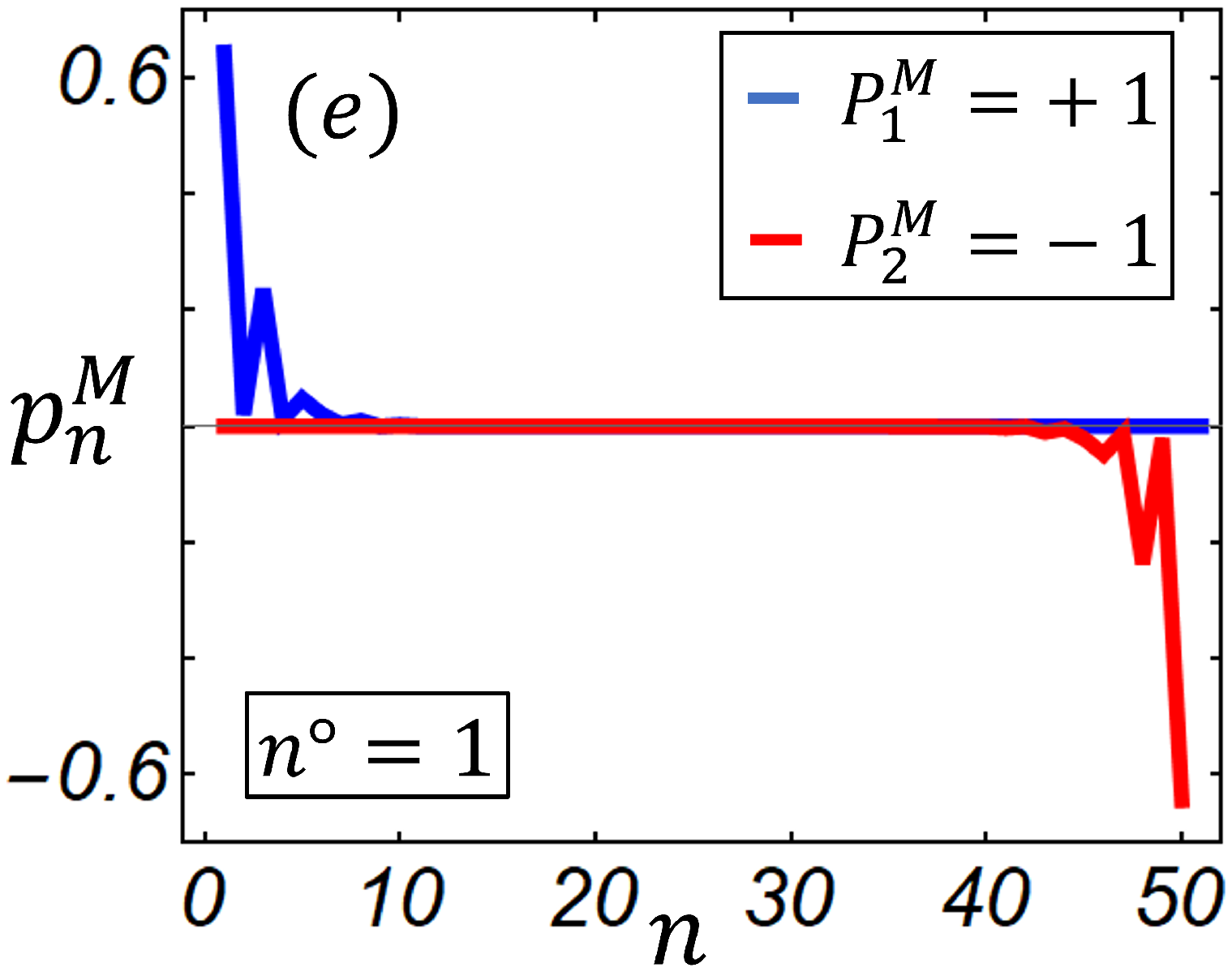}
	\hspace{0.05cm}
	\includegraphics[scale=0.27]{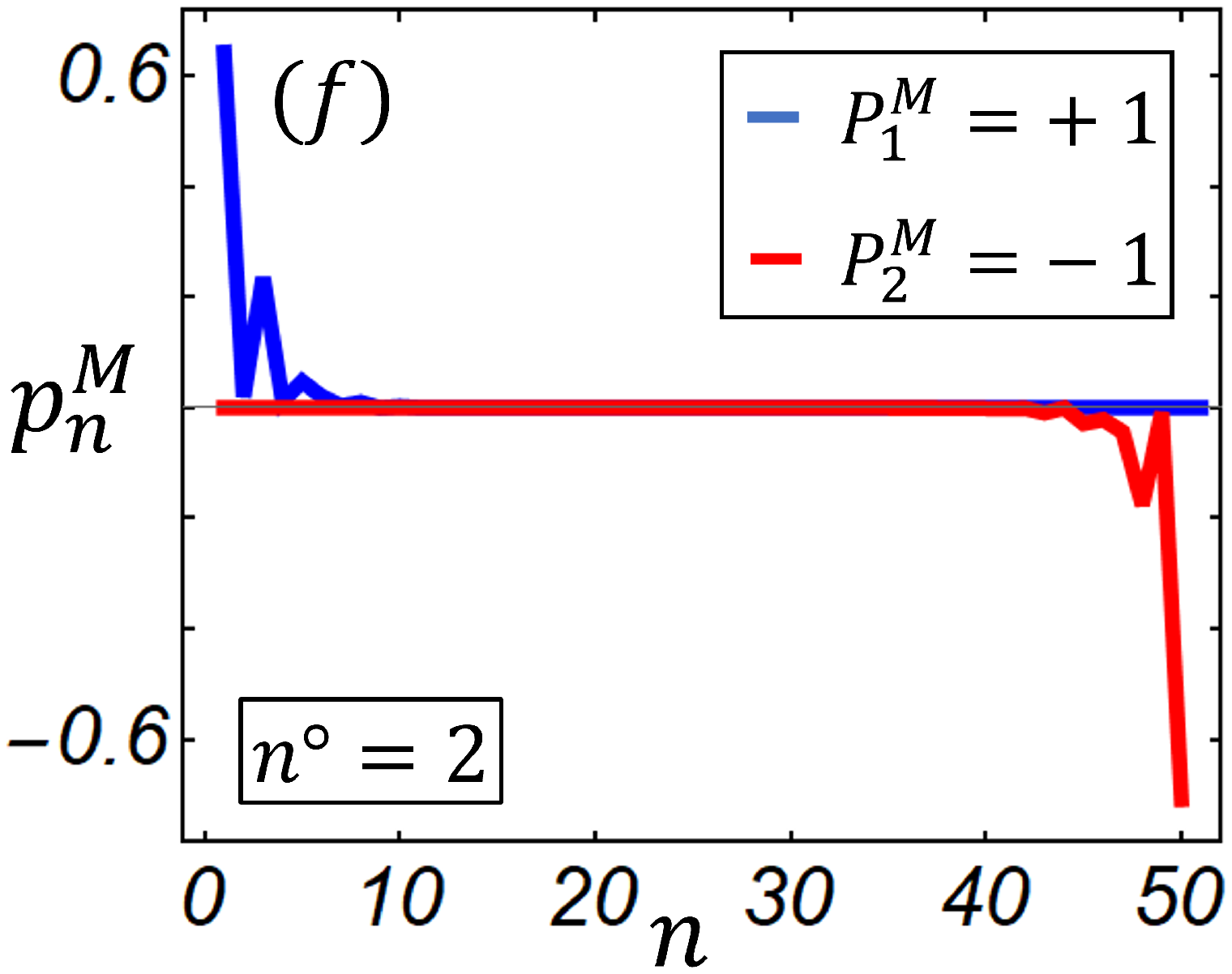}
	\hspace{0.05cm}
	\includegraphics[scale=0.27]{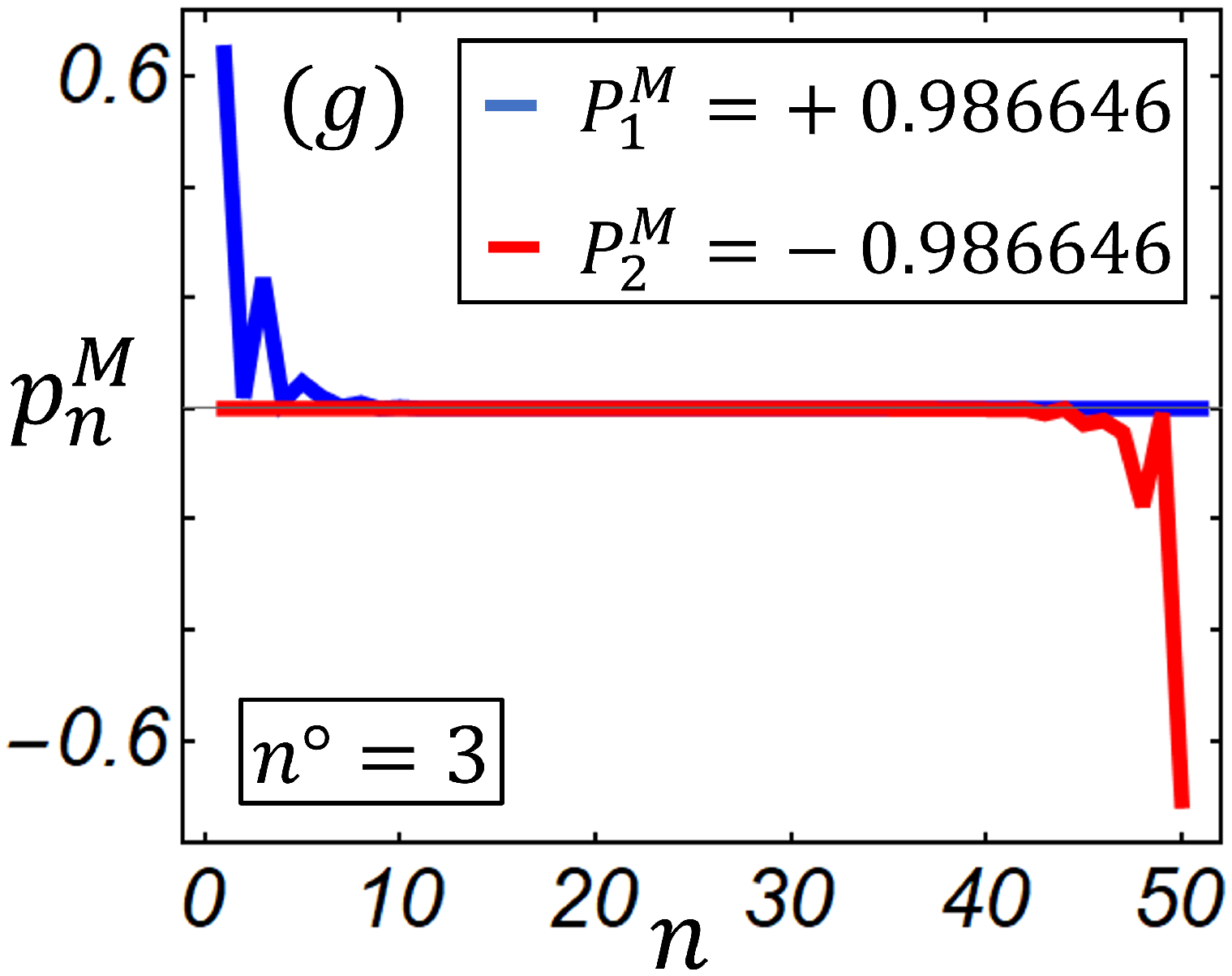}
	\hspace{0.05cm}
	\includegraphics[scale=0.27]{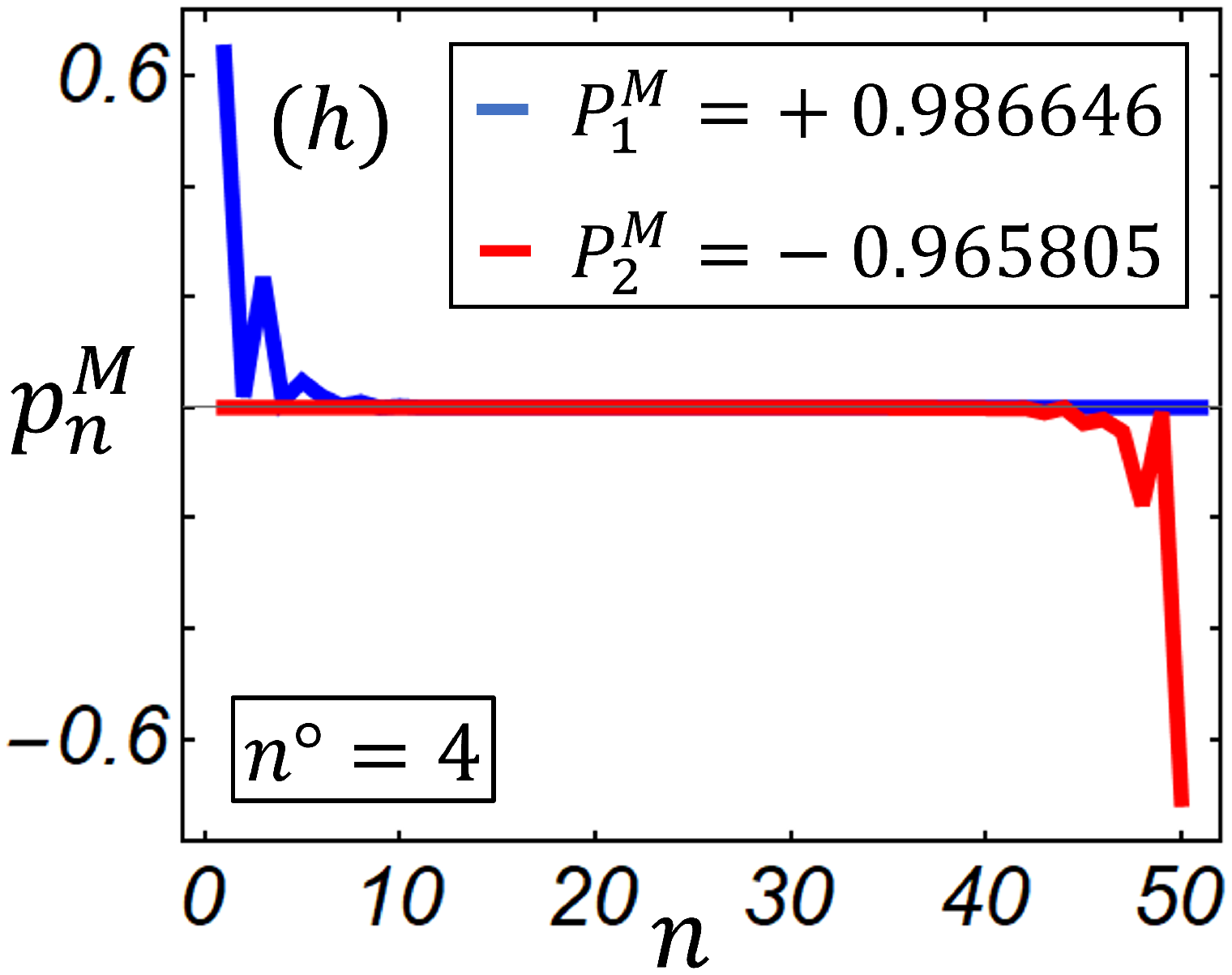}
	\caption{Panels (a) to (d): Site-dependent charge density of the sub-gap modes. The insets in panels (a) and (b) show details of the corresponding behaviors. Panels (e) to (h): Local Majorana polarization of the same modes. The total charge densities and the total Majorana polarizations are reported, respectively, in the insets of panels (a) to (d) and of panels (e) to (h). Each panel refers to the boundary conditions reported in table (d) of Fig. \ref{Scheme}. The remaining parameters have been fixed as: $\mu=0.5$, $t=1$, $\Delta=0.3$, and $L=51$.}
	\label{DensityChargePolarization}
\end{figure*}

The degree of hybridization can be quantified by studying the site-dependent charge density, $q_n=|u_n|^2-|v_n|^2$, as reported in Fig. \ref{DensityChargePolarization} (a)-(d). Majorana modes are neutral sub-gap excitations and thus deviation from this condition provides indications about the degree of hybridization. The charge neutrality condition is guaranteed when particle-hole symmetry is preserved, see panels (a) and (b) of Fig. \ref{DensityChargePolarization}, so that the total charge density $Q=\sum_{n=1}^{L}q_n$ vanishes for both sub-gap modes.

When particle-hole symmetry is broken, see panels (c) and (d) of Fig. \ref{DensityChargePolarization}, a non-vanishing charge appears at the chain boundaries; in particular, the negative sign of the total charge density $Q$ provides evidence of the hole-like hybridization of the edge modes. These charged modes originate from strongly hybridized Majorana excitations which have lost their topological protection due to the breaking of the particle-hole symmetry. The type of hybridization, either electron-like or hole-like, or equivalently the sign of the charge acquired by the end modes, is hardly predictable. It is an emerging property of the entire system which depends on the spectral properties and on the specific set of the chain parameters, as discussed in detail in Appendix \ref{AppA}.

Degradation of the topological properties can be also quantified by using the total Majorana polarization ($P^M$) \cite{PhysRevLett.108.096802,PhysRevLett.110.087001,BENA2017349,PhysRevB.92.115115,MaiellaroGeoFrust}, which measures the quasiparticles weight in Nambu space. The total Majorana polarization can be written in terms of the site-dependent Majorana polarizations $p^M_n$ \cite{PhysRevLett.108.096802,BENA2017349} according to the expression $P^M=\sum_{n=1}^{L}p^M_n$. Genuine Majorana modes, labelled by $j \in \{1,\ 2\}$, are characterized by opposite polarization values, namely $P^M_1=1$ and $P^M_2=-1$, while $p^M_n$ is a peaked function at the system edges. The above features originate from the impossibility to isolate a Majorana monopole.

The site-dependent Majorana polarizations $p^M_n$ and the Majorana polarizations $P^M_j$ of the sub-gap modes are reported in panels (e) to (h) of Fig. \ref{DensityChargePolarization}. Systems with particle-hole symmetry and generalized boundary conditions, see panels (e)-(f), feature sub-gap modes with polarization values comparable to those of the Kitaev chain with OBCs. When the particle-hole symmetry is broken, see panels (g)-(h), the Majorana polarization is lowered and we observe a net loss: $|P^{M}_{1,2}|<1$. Moreover, the site-dependence of $p^M_n$ is not heavily perturbed by the generalized boundary conditions. At a close inspection of panels (e)-(g), we observe that the relation $P^M_1+P^M_2=0$ is invariably respected. When both reflection and particle-hole symmetries are broken, see panel (h) of Fig. \ref{DensityChargePolarization}, a residual topological charge $P^M_1+P^M_2 \sim 0.02$ is induced. Majorana polarization losses and residual topological charges are signatures of the hybridization of the Kitaev chain with the environmental degrees of freedom and cannot be observed in isolated systems such as a Kitaev chain with OBCs.

\begin{figure*}
	\includegraphics[scale=0.45]{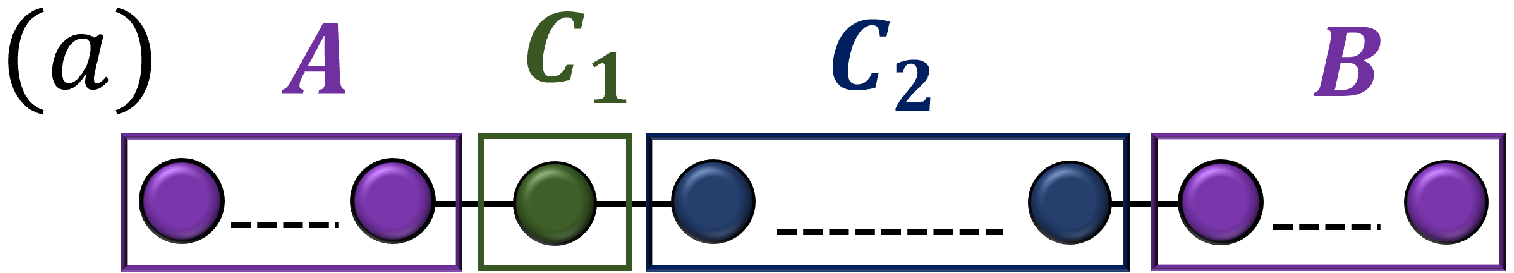}\\
	\vspace{0.5cm}
	\includegraphics[scale=0.4]{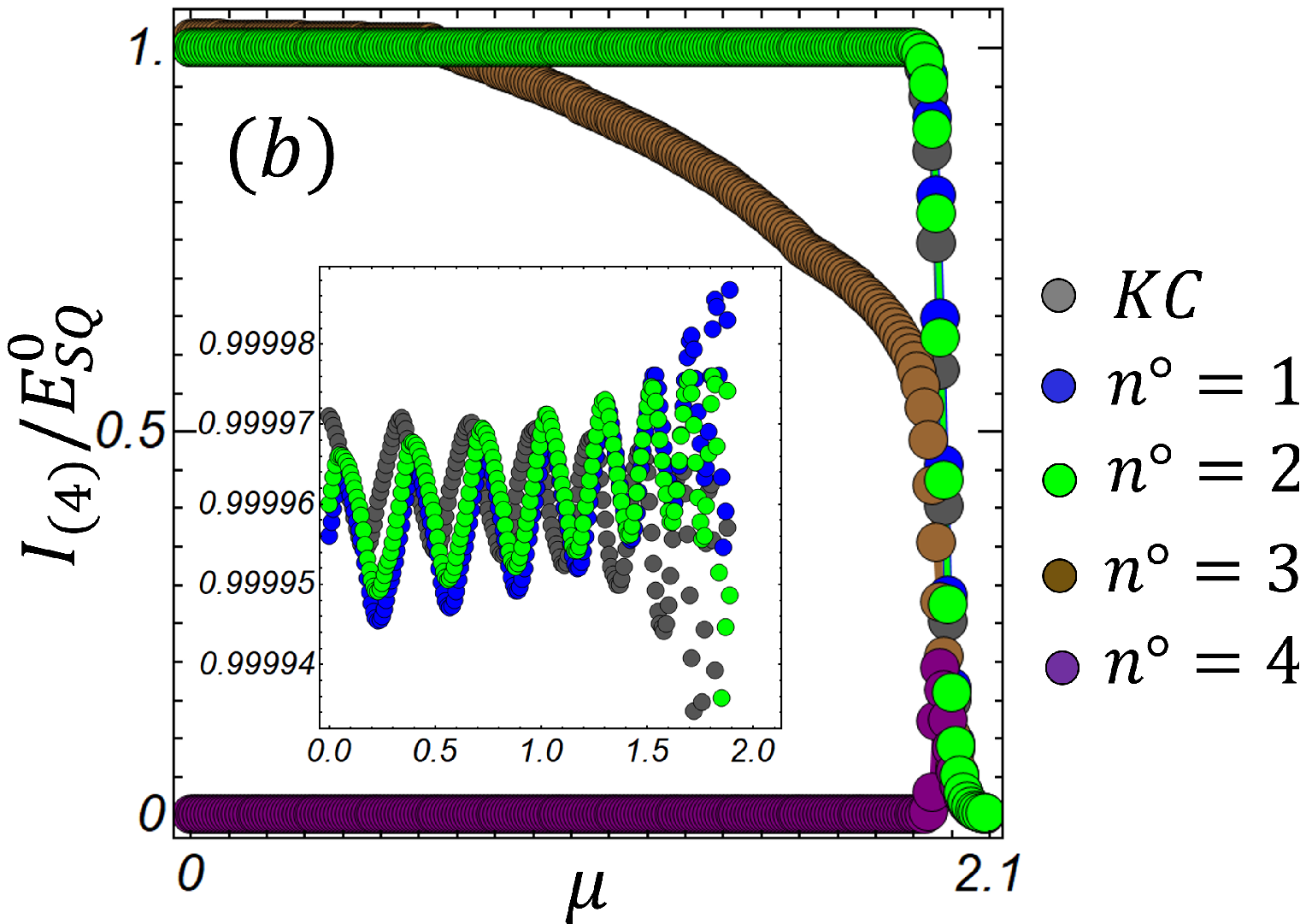}
	\hspace{0.1cm}
	\includegraphics[scale=0.32]{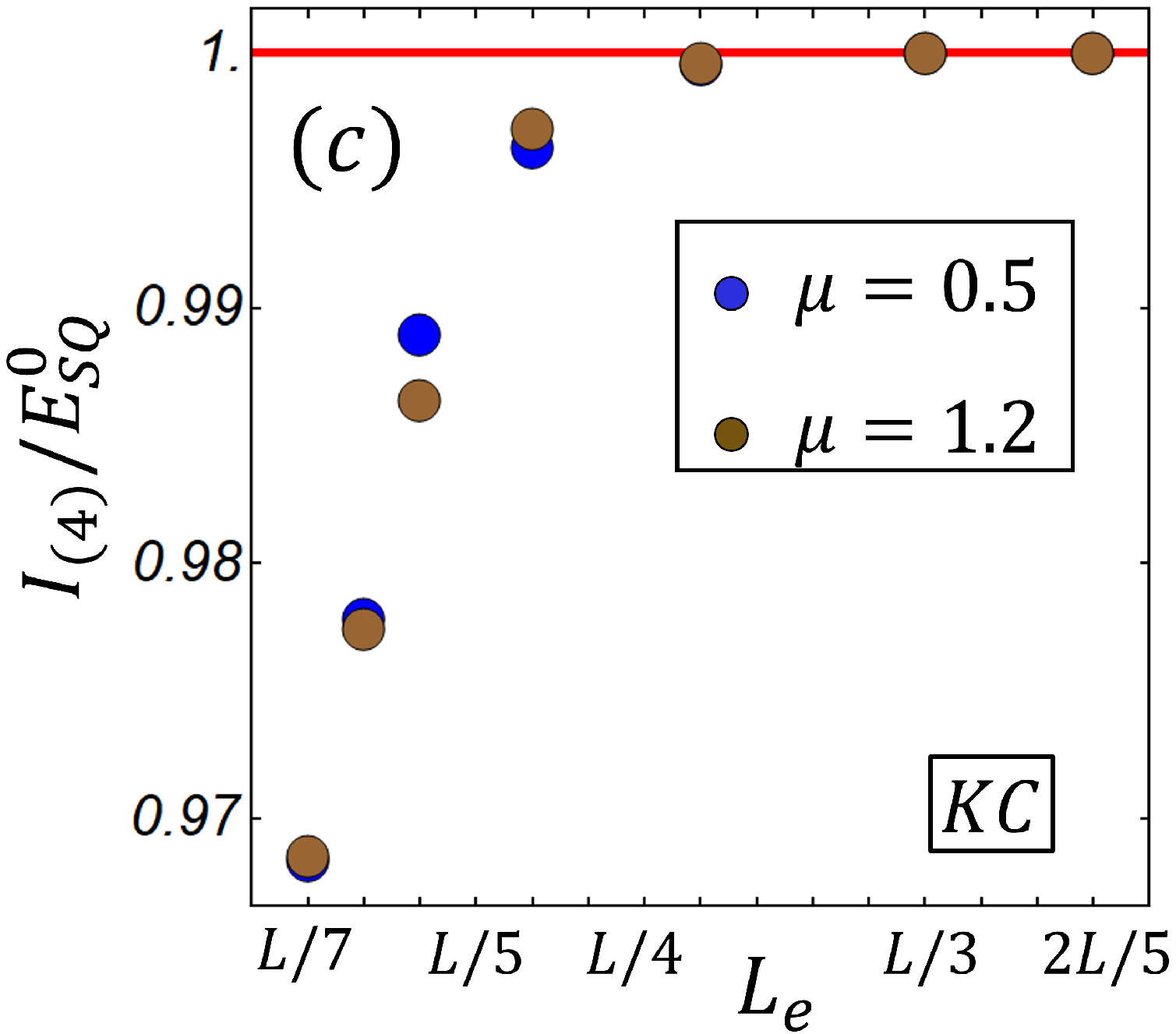}
	\includegraphics[scale=0.3]{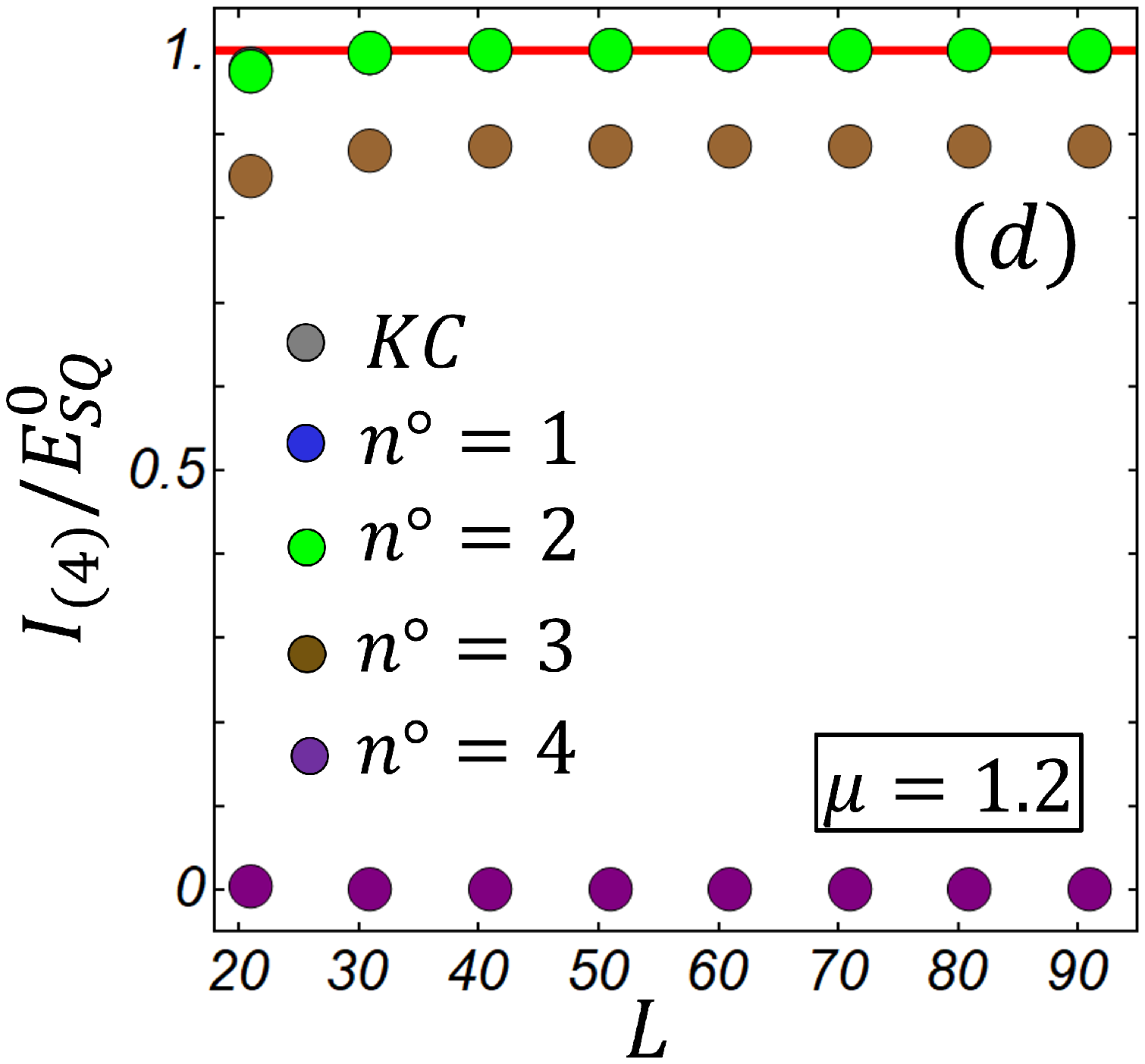}
	\caption{Panel (a): Sketch of the chain quadripartition exploited to compute the edge-to-edge quantum conditional mutual information $I_{(4)}$ and the quantized unit of topological squashed entanglement $E_{SQ}^0$ between the edges. The bulk is partitioned in two subsystems $C_1$ and $C_2$ with $L_{C_{1}}=1$, while the two remaining parts of the chain are the edges, denoted by $A$ and $B$. Panel (b): Behavior of the ratio $I_{(4)}/E^0_{SQ}$ as a function of the chemical potential $\mu$. The different curves correspond to the different choices of generalized boundary conditions reported in table (d) of Fig. \ref{Scheme}. The size $L$ of the chain and the size $L_e$ of the edges have been fixed, respectively, at $L=51$ and $L_e = L/3 = 17$. The inset magnifies the differences between the Kitaev chain ($KC$) with OBCs and the cases $n\degree=1$ and $n \degree=2$. Panel (c): Behavior of the ratio $I_{(4)}/E^0_{SQ}$ as a function of the boundary length $L_e$. Panel (d): Behavior of the ratio $I_{(4)}/E^0_{SQ}$ as a function of the chain length $L$ (color codes as in panel (b)). The red horizontal lines in panels (c) and (d) mark the maximum quantized value $\log{2}/2$ that is reached by $E^0_{SQ}$ at the exact ground-state topological degeneracy point, $\mu = 0$, for a Kitaev chain with OBCs hosting genuine Majorana modes. In all panels the Hamiltonian hopping and pairing amplitudes are fixed at $t=1$ and $\Delta=0.3$.}
	\label{Entanglement}
\end{figure*}

So far we have discussed important features of topological sub-gap states. Some of the mentioned features however are not always necessarily associated with Majorana modes. For instance, zero-energy sub-gap excitations can emerge in the physics of Andreev bound states \cite{AguadoABS,PhysRevB.96.075161,PhysRevB.97.214502,PhysRevB.96.075161,PhysRevLett.104.067001}. Moreover, a trivial superconducting phase can host charge-neutral quasiparticles \cite{PhysRevLett.116.246403,Denisov_2021,Sato_2019}. These observations show the need to identify unequivocal signatures of topological order. For systems where the bulk-edge correspondence applies, an unambiguous signature of a topologically ordered phase is provided by topological invariants \cite{PhysRevB.55.1142} which count the number of inversions between valence and conduction bands \cite{PhysRevB.103.075139}. In the absence of translational invariance and whenever the bulk-edge correspondence does not hold, alternative approaches are needed to identify the topological nature of the edge modes.

One such approach relies on the determination of the unique nonlocal quantum correlations that are established in a topologically ordered phase between the system edges. Indeed, such topological nonlocal edge-to-edge correlations are faithfully quantified by a specific measure of bipartite entanglement \cite{doi:10.1063/1.1643788,5075874}, the squashed entanglement (SE) $E_{SQ}^0$ between the edges, obtained by a suitable quadripartition of the system, and by its natural upper bound, the edge-to-edge quantum conditional mutual information (QCMI) $I_{(4)}$, as recently shown in Refs. \cite{MaieIllum1,MaieIllum2}. The edge-edge QCMI provides an upper bound on the long-distance, edge-edge squashed entanglement obtained by four-partitioning the system into the two edges and a bipartition of the bulk. The QCMI upper bound on the SE is then obtained by a suitable combination of reduced von Neumann entropies stemming from the quadripartition; this combination squashes out the classical contributions leaving only the genuinely quantum ones.
Being the topological order encoded in the edges, the edge-edge SE $E^0_{SQ}$ identifies unequivocally the  topologically ordered phases, also fulfilling all the criteria of a genuine nonlocal order parameter \cite{MaieIllum1,MaieIllum2}. In particular, the edge-edge SE assumes the quantized value $E^0_{SQ} = \log{2}/2$, i.e. half of the maximal Bell-pair entanglement, at the exact ground-state topological degeneracy point, $\mu = 0$, for a Kitaev chain with OBCs hosting genuine Majorana modes, and remains constant at this quantized value throughout the entire topologically ordered phase, up to $\mu = 2t$.

The edge-edge QCMI $I_{(4)}$ realizing the natural upper bound on $E^0_{SQ}$ is expressed in terms of the reduced von Neumann entropies as follows:
\begin{equation}
I_{(4)} = S_{AC_1} + S_{BC_1} - S_{ABC_1} - S_{C_1} \, ,
\end{equation}
where $S(XY)$ denotes the von Neumann entropy computed on the many-body ground state of the reduced system running on sites belonging to the parts $X$ and $Y$. The four different labels $A$, $B$, $C_1$, and $C_2$ identify the four parts in which the chain is quadripartite. $A$ and $B$ refer to the two edges, while the bulk is bipartite into two parts $C_1$ and $C_2$. The latter is the part of the bulk which is traced out in the beginning to leave the reduced ground state of the three remaining parts, so it does not appear in the definition of $I_{(4)}$, while the remaining part of the bulk $C_1$ separates the two edges. The true edge-edge SE is then obtained taking the infimum of $I_{(4)}$ over all possible states of $ABC_1$. Both the edge-edge SE and the edge-edge QCMI are insensitive to the relative lengths of the two parts of the bulk and thus, without loss of generality, we can set $L_{C_1}=1$, while also taking symmetric edges $L_A = L_B = L_e$. Details on the system partition are illustrated in panel (a) of Fig. \ref{Entanglement}.

In panel (b) of Fig. \ref{Entanglement} we report the behavior of the ratio between the edge-edge QCMI $I_{(4)}$ and the maximum quantized value of the edge-edge SE $E^0_{SQ}$ as a function of the chemical potential $\mu$, for boundary conditions belonging to the distinct symmetry classes listed in table (d) of Fig. \ref{Scheme}. The behavior of $I_{(4)}/E^0_{SQ}$ as a function of $\mu$ for a Kitaev chain with OBCs is also reported for comparison. The plots show that the $I_{(4)}$ \textit{vs} $\mu$ curves closely match the curve corresponding to a Kitaev chain with OBCs as long as the particle-hole symmetry is preserved. When this symmetry is broken (brown and purple curves), the $I_{(4)}$ \textit{vs} $\mu$ curves feature a significant deviation from the step-jump behavior characterizing the particle-hole symmetric case. In particular, the simultaneous breaking of both the particle-hole and reflection symmetry (purple plot) induces a complete degradation of the topological properties which is signaled by the vanishing of $I_{(4)}$ for $\mu \lesssim 2t$. In the close neighborhood of $\mu \approx 2t$, we observe a revival of the nonlocal correlation, signaled by a peak in the $I_{(4)}$ vs $\mu$ curve; this phenomenon is induced by the mitigation effects of the gap closure anticipating the transition to the topologically trivial phase.

We have carried out the study reported in panel (b) of Fig. \ref{Entanglement} by setting the length of the boundary subsystems $A$ and $B$ at $L_e=L/3$; it is of course possible to show that indeed the results are not affected by the specific partition strategy. In panel (c) of Fig. \ref{Entanglement} we analyze the behavior of the ratio $I_{(4)}/E^0_{SQ}$ as a function of $L_e$ for a Kitaev chain with OBCs of length $L=51$, which is used as reference system to compare the different partition strategies. The aforementioned curves show that the edge-edge QCMI $I_{(4)}$ reaches the quantized value $E^0_{SQ} = \log{2}/2$, expected for a Kitaev chain hosting genuine Majorana modes, exactly at $L_e=L/3$. With the mentioned choice for the partition procedure, the non-local correlations captured by $I_{(4)}$ and $E^0_{SQ}$ are not affected by finite-size effects, as shown in panel (d) of Fig. \ref{Entanglement}, proving the robustness of the edge-edge entanglement.

The results reported in panel (b) of Fig. \ref{Entanglement} clearly imply that, as long as $\mu<0.7t$, nonlocal correlations of a Kitaev chain in the symmetry class $\overline{PH}/R$ are comparable to those of a standard Kitaev chain with OBCs. On the other hand, the simultaneous breaking of the particle-hole and the inversion symmetry (symmetry class $\overline{PH}/\overline{R}$) originates a complete degradation of the topological nonlocal correlations, as signaled by the vanishing of the edge-edge QCMI and thus of the edge-edge SE.

\section{conclusions and outlook}
\label{conclusions}
We have studied the topological properties of a Kitaev chain under the perturbing influence of generic boundary conditions mimicking the coupling with external degrees of freedom (e.g. external reservoirs). Using appropriate constraints, we have derived a four-parameter, symmetry-based classification of the different possible boundary effects showing that particle-hole and reflection symmetries can be broken or preserved by appropriately fixing the boundary parameters. When the particle-hole and the reflection symmetries are simultaneously broken, the topological protection of the edge modes is completely lost due to the hybridization with the external degrees of freedom. Vice-versa, the edge modes turn out to be quite robust in the intermediate regimes. We have investigated the resilience of topological states by using several estimators, also including the recently introduced edge-edge quantum conditional mutual information and the related edge-edge squashed entanglement that provides a faithful and \emph{bona fide} measure of the nonlocal quantum correlations between the Majorana excitations. We have shown that these information and entanglement quantifiers, complemented by the energy eigenvalues and the charge densities of the edge modes, yield a complete characterization of the topological properties of the system and of their fate in the different ranges of boundary conditions.

In a future perspective, we plan to apply edge-edge mutual information and squashed entanglement to the investigation of the open dynamics beyond the stationarity and vanishing currents regimes, in order to characterize the fate of topological order in generic non-equilibrium dynamics, possibly modeled either by means of effective non-Hermitian Hamiltonians or in the full generality of completely positive and trace preserving maps.

{\it Acknowledgments} -- We acknowledge support by MUR (Ministero dell’Università e della Ricerca) via the project PRIN 2017 ”Taming complexity via QUantum Strategies: a Hybrid Integrated Photonic approach” (QUSHIP) Id. 2017SRNBRK.

\appendix

\section{Charge density and spectral properties}
\label{AppA}

Despite the lack of an immediate correspondence between the charge $Q$ and the energy $E$ of a normalized eigenstate of the Kitaev chain Hamiltonian, the two quantities are not independent. This statement is easily proven by using the stationary Bogoliubov-de Gennes equations (\ref{BdGKC}) complemented by the normalized expression of the charge density $Q=\sum_{n=1}^{L} \psi^\dagger_n \sigma_z \psi_n$. For a Kitaev chain with generalized boundary conditions, it is rather straightforward to prove the validity of the relation
\begin{eqnarray}
	\begin{split}
	&\mu Q=2 Re\biggl[\sum_{n=1}^{L-1}\psi_n^\dagger  \left(
		\begin{array}{cc}
		-t&\Delta\\
			-\Delta&t\\
		\end{array}
		\right)
		\psi_{n+1}\biggr]-E+\\
		&\frac{\Delta^2-t^2}{t} \biggr[ a_l|u_1|^2+a_r|u_L|^2-\bigr(d_l|v_1|^2+d_r|v_L|^2\bigl)\biggl] .
		\end{split}
	\label{ChargeGeneral}
\end{eqnarray}

When one considers a Kitaev chain with OBCs (i.e. $a_{l,r}=d_{l,r}=0$), Eq. (\ref{ChargeGeneral}) takes the following form
\begin{eqnarray}
		&\mu Q=2 Re\biggl[\sum_{n=1}^{L-1}\psi_n^\dagger  \left(
		\begin{array}{cc}
			-t&\Delta\\
			-\Delta&t\\
		\end{array}
		\right)
		\psi_{n+1}\biggr]-E.
		\label{ChargeKC}
\end{eqnarray}

A Kitaev chain with OBCs is a particle-hole symmetric model implying that any eigenstate $\psi_n$ of the model Hamiltonian with energy eigenvalue $E$ comes together with a particle-hole symmetric state $\tilde{\psi}_n=P \psi_n$ with energy eigenvalue $\tilde{E}=-E$. This correspondence is implemented mathematically by the particle-hole operator $P=\sigma_x K$ written in terms of the conjugation operator $K$. Using Eq. (\ref{ChargeKC}) one can show that $\psi_n$ and $\tilde{\psi}_n$ have opposite charge (i.e. $\tilde{Q}=-Q$). A few lines proof is obtained by observing that
\begin{eqnarray}
	\begin{split}
	&\mu \tilde{Q}=2 Re\biggl[\sum_{n=1}^{L-1}\tilde{\psi}_n^\dagger  \left(
\begin{array}{cc}
	-t&\Delta\\
	-\Delta&t\\
\end{array}
\right)
\tilde{\psi}_{n+1}\biggr]-\tilde{E}=\\
&-\biggl\{2 Re\biggl[\sum_{n=1}^{L-1}\psi_n^\dagger  \left(
\begin{array}{cc}
	-t&\Delta\\
	-\Delta&t\\
\end{array}
\right)
\psi_{n+1}\biggr]-E \biggr\}=-\mu Q,
	\end{split}
\label{PHCharge}
\end{eqnarray}
where the relation $\sigma_x (-t \sigma_z+i \Delta \sigma_y)\sigma_x=-(-t \sigma_z+i \Delta \sigma_y)$ has been used in the derivation.

\begin{figure*}
	\includegraphics[scale=0.27]{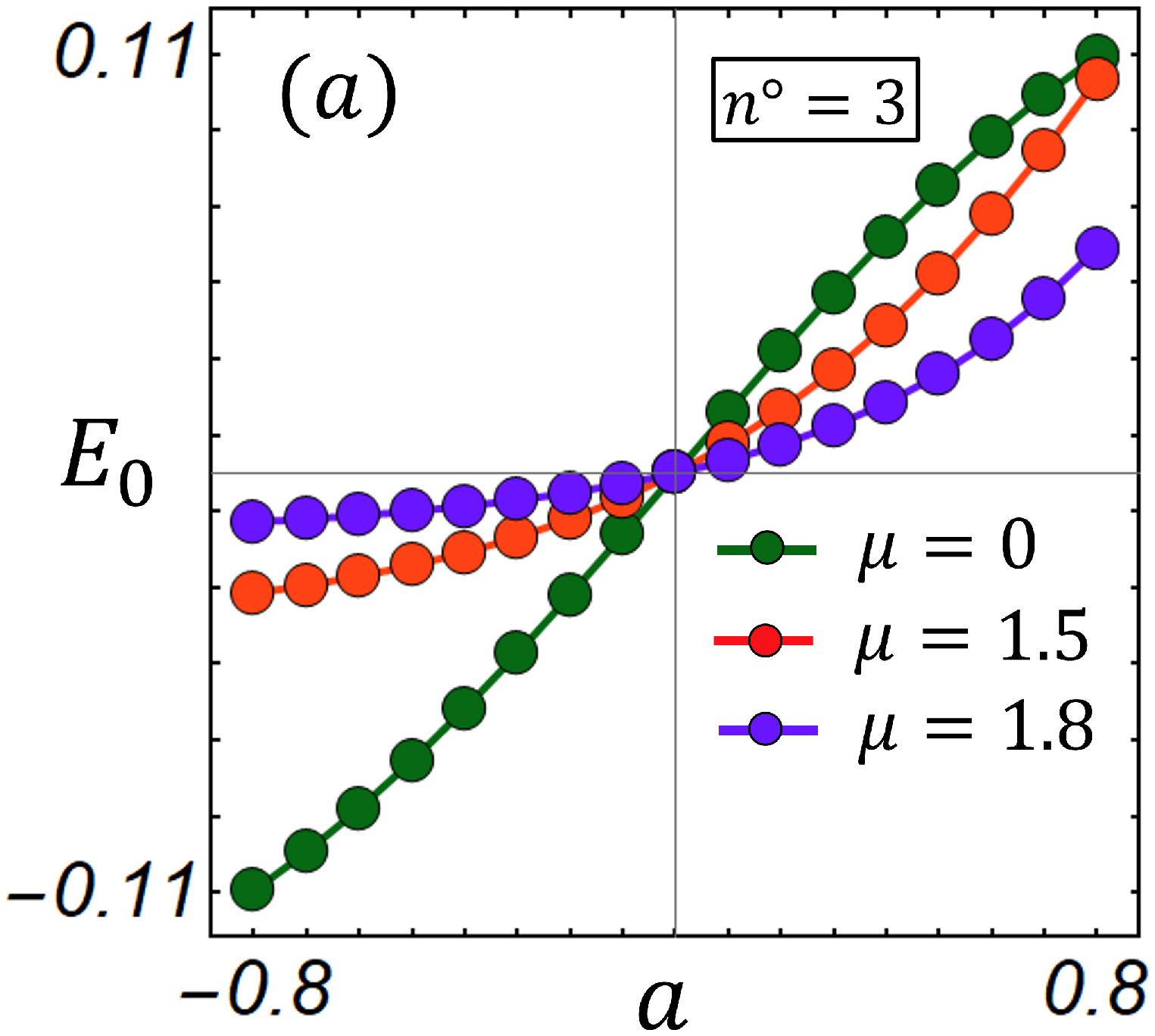}
	\hspace{0.1cm}
	\includegraphics[scale=0.27]{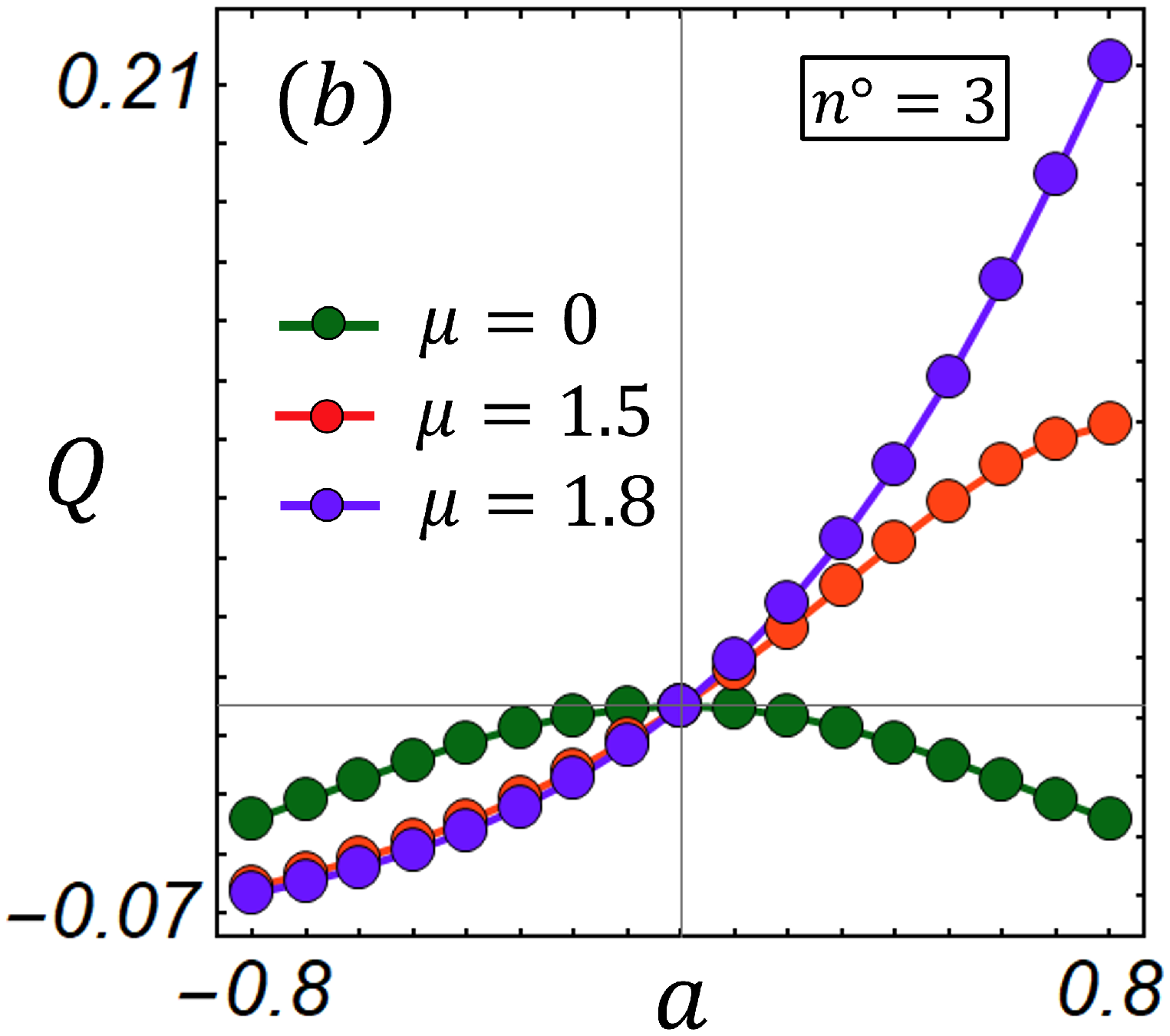}
		\hspace{0.1cm}
	\includegraphics[scale=0.27]{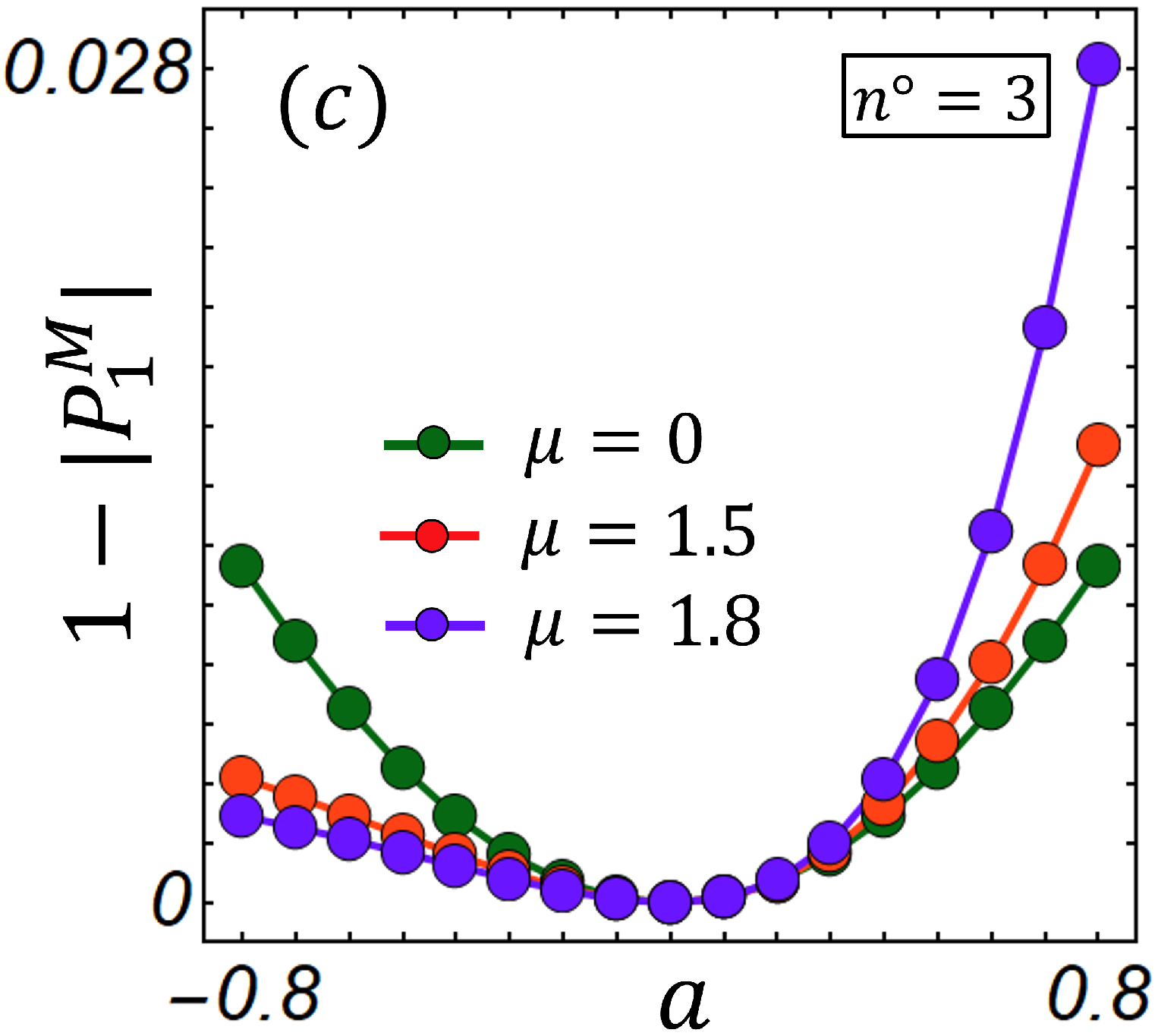}
	\includegraphics[scale=0.282]{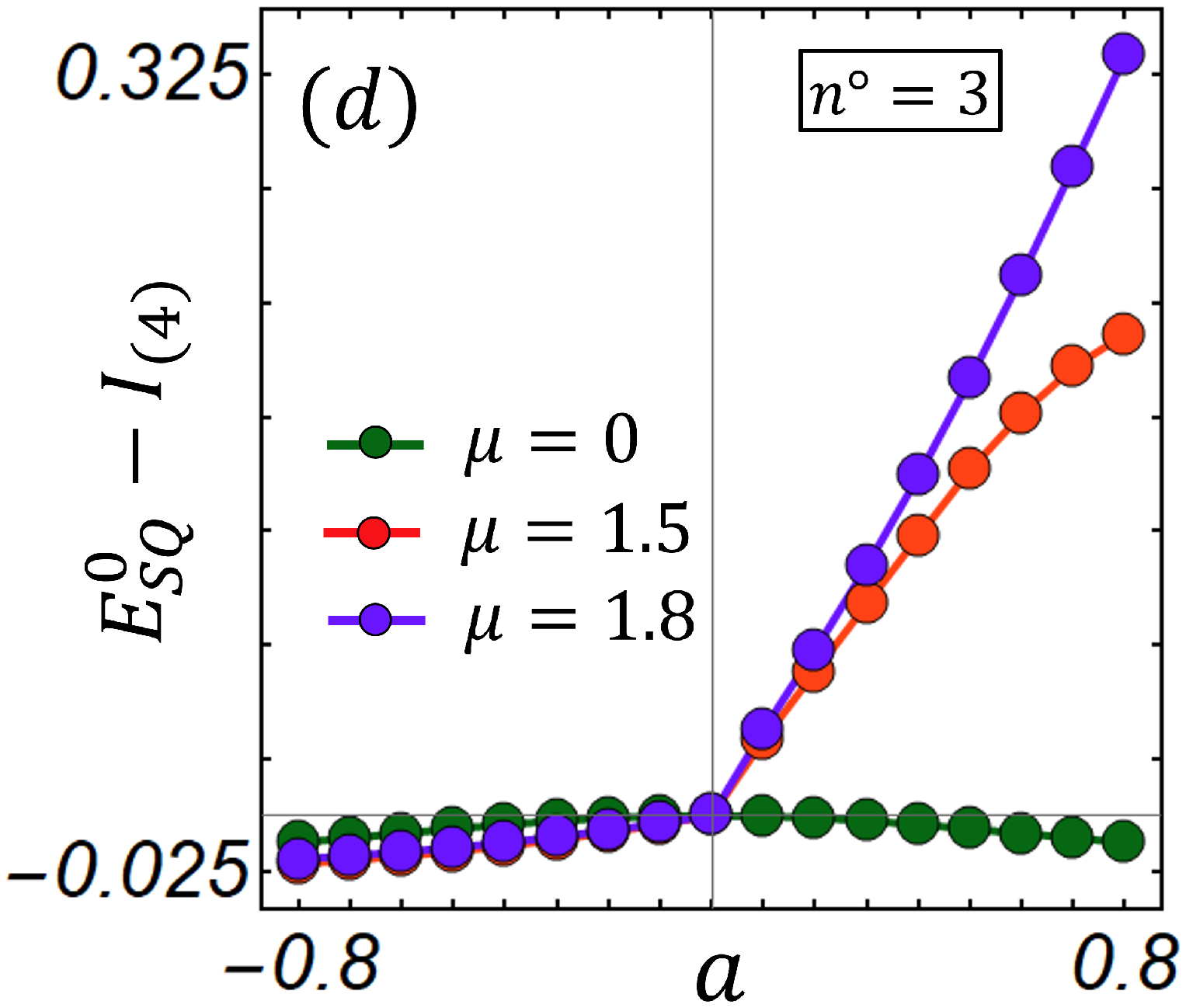}
	\caption{Panel (a): Sub-gap energy level $E_0$ as a function of the boundary parameter $a$. Panel (b): Charge excess $Q$ as a function of $a$. Panel (c): Polarization loss $1-|P^{M}_1|$ as a function of $a$. Panel (d): Difference between the quantized value of the edge SE $E^0_{SQ} = \log{2}/2$ and the edge QCMI $I_{(4)}$. Different curves correspond to different values of the chemical potential, as specified in the figure insets. Throughout, we have considered the symmetry class $\overline{PH}/R$. In all panels the Hamiltonian hopping and pairing amplitudes are fixed at $t=1$ and $\Delta=0.3$.}
	\label{vsa}
\end{figure*}

The above observations imply that as long as the particle-hole symmetry is preserved the total electric charge associated to states with positive energy is completely balanced by the opposite charge related to states belonging to the negative part of the energy spectrum.

Interestingly, despite the fact that particle-hole symmetry breaking induces a violation of the charge neutrality of the sub-gap modes, charge's excess/deficiency of these states is completely neutralized by the electric charge of the remaining bulk-like states. A similar neutralization mechanism also operates for the Majorana polarization loss discussed in Appendix \ref{AppB} [\onlinecite{PhysRevB.85.235307}].

\section{Sub-gap energy levels, charge excess, polarization loss, and edge entanglement}
\label{AppB}

In this appendix we review the behavior of the different topological estimators as functions of the boundary parameter $a$, for different values of the chemical potential. In panel (a) of Fig. \ref{vsa} we report the behavior of the sub-gap energy levels; in panel (b) the charge excess; in panel (c) the polarization loss; and in panel (d) the difference between the quantized value of the edge-edge SE $E^0_{SQ} = \log{2}/2$ and the edge-edge QCMI $I_{(4)}$. To be definite, we consider the symmetry class $\overline{PH}/R$, while a similar analysis can be performed for systems belonging to the symmetry class $\overline{PH}/\overline{R}$.

The plots in Fig. \ref{vsa} (a) show that the sub-gap energy levels features a nonlinear dependence on the boundary parameter $a$ that nevertheless is monotonically increasing along the entire range of variations of $a$ for any value of the chemical potential.

Fig. \ref{vsa} (b) shows the complicated behavior of the charge excess. As anticipated earlier, the sign of this quantity is not directly related to the sign of the boundary parameter; it is significantly affected by the actual value of the chemical potential.

Fig. \ref{vsa} (c) indicates that despite the significant range of variations of the boundary parameter $a$ that we have considered, the polarization loss never exceeds a few percent of the Majorana polarization of sub-gap states for a Kitaev chain with OBCs.

Finally, Fig. \ref{vsa} (d) shows that for all negative values of the $a$ parameter the edge-edge QCMI $I_{(4)}$ is strongly resilient and remains very close to the quantized value $E^0_{SQ} = \log{2}/2$ that the edge-edge SE takes in the topologically ordered phase ($\mu \leq 2t$) of a Kitaev chain with OBCs. In the interval of positive values of the boundary parameter, the QCMI $I_{(4)}$ remains completely resilient at the point of exact topological degeneracy $\mu = 0$, while for larger values of the chemical potential it starts to deviate significantly from $E^0_{SQ}$.

\bibliography{Bibtex_KCBoundary}
%\bibitem{PHSb}F. Setiawan and Jay D. Sau, Electron-boson-interaction induced particle-hole symmetry breaking of conductance into subgap states in superconductors, Physical Review Research \textbf{3}, L032038 (2021).
\end{document}